\documentclass[preprint,aps,prl,superscriptaddress]{revtex4-1}

\usepackage{graphicx}  
\usepackage{dcolumn}   
\usepackage{bm}        
\usepackage{amssymb}   
\usepackage{amsmath}   
\usepackage{mathtools} 
\usepackage{tabularx}  
\usepackage{lipsum}    

\begin{document}

\title{\bf{Collective motion of driven semiflexible filaments tuned by soft repulsion and stiffness}} 
\author{Jeffrey M. Moore}
\author{Tyler N. Thompson}
\author{Matthew A. Glaser} 


\affiliation{Department of Physics, University of Colorado, Boulder, CO 80309}

\author{Meredith D. Betterton}
\affiliation{Department of Physics, University of Colorado, Boulder, CO 80309}
\affiliation{Department of Molecular, Cellular, and Developmental Biology, University of Colorado, Boulder, CO 80309}

\date{\today}

\begin{abstract}
  
In active matter systems, self-propelled particles can self-organize to undergo collective motion, leading to persistent dynamical behavior out of equilibrium. In cells, cytoskeletal filaments and motor proteins self-organize into complex structures important for cell mechanics, motility, and division. Collective dynamics of cytoskeletal systems can be reconstituted using filament gliding experiments, in which cytoskeletal filaments are propelled by surface-bound motor proteins. These experiments have observed diverse dynamical states, including flocks, polar streams, swirling vortices, and single-filament spirals. Recent experiments with microtubules and kinesin motor proteins found that the collective behavior of gliding filaments can be tuned by altering the concentration of the crowding macromolecule methylcellulose in solution. Increasing the methycellulose concentration reduced filament crossing, promoted alignment, and led to a transition from active, isotropically oriented filaments to locally aligned polar streams. This emergence of collective motion is typically explained as an increase in alignment interactions by Vicsek-type models of active polar particles. However, it is not yet understood how steric interactions and bending stiffness modify the collective behavior of active semiflexible filaments. Here we use simulations of driven filaments with tunable soft repulsion and rigidity in order to better understand how the interplay between filament flexibility and steric effects can lead to different active dynamic states. We find that increasing filament stiffness decreases the probability of filament alignment, yet increases collective motion and long-range order, in contrast to the assumptions of a Vicsek-type model. We identify swirling flocks, polar streams, buckling bands, and spirals, and describe the physics that govern transitions between these states. In addition to repulsion and driving, tuning filament stiffness can promote collective behavior, and controls the transition between active isotropic filaments, locally aligned flocks, and polar streams.

\end{abstract}

\maketitle



Active particles exhibit complex and dynamical order at length scales much larger than the scale of a single particle.  Living systems with collective dynamics include swimming bacteria, schools of fish, flocks of birds, and crowds of people~\cite{vicsek12, czirok96, dombrowski04, parrish02, bajec09, silverberg13a}. The study of active biopolymers is motivated by the activity of cells, because intracellular organization and dynamics are largely governed by the cytoskeleton.  Cytoskeletal filaments and motor proteins generate active forces that control the assembly of critical cellular structures and long-range patterns~\cite{hyman96, nedelec97, surrey01}, or active nematics~\cite{sanchez12, m.lemma19}.  Collective behavior can also occur in filament gliding experiments, wherein cytoskeletal filaments are propelled by motor proteins bound to a surface. Previous work on filament gliding has reported several nonequilibrium dynamical states~\cite{butt10, schaller10, schaller11, liu11, sumino12, huber18}. However, our understanding of the physics that controls these phases is incomplete.

Systems of microtubules propelled by kinesin motors often do not interact and align sufficiently for the emergence of collective motion~\cite{liu11, sumino12, inoue15}. Recent work showed that adding methylcellulose as a molecular crowding agent can reduce filament crossing and cause microtubules to locally align~\cite{inoue15, saito17, kim18}. This has been proposed to occur due to attractive depletion forces that lead filaments to align and form polar bundles~\cite{asakura58, marenduzzo06, lau09, sanchez12, inoue15, saito17, kim18, huber18, m.lemma19}. Such systems may be well-described by Vicsek-type models of active polar particles, wherein collective motion is governed by local alignment interactions~\cite{vicsek95, levine00, gregoire04, aldana07, chate08}. However, previous rheological work reported that depletion forces alone were insufficient to explain the degree of bundling observed in systems of actin and methylcellulose, and proposed that filaments might be kinetically trapped due to the formation of a methycellulose network~\cite{kohler08}. In addition, microtubule--kinesin gliding experiments observed the alignment of filaments due to local steric interactions in the absence of any depletants, including the formation of self-interacting, single-filament spirals~\cite{liu11, inoue15, saito17}. Therefore, we seek to understand the phenomenology observed in microtubule--kinesin filament gliding experiments with methylcellulose, in the context of purely repulsive filament interactions that are tunable by the methylcellulose concentration. Most previous theoretical and computational work investigating collective behavior of active filaments with repulsive interactions focused on purely rigid rods~\cite{kraikivski06, peruani06, baskaran08, ginelli10, yang10, peruani11, peruani12, abkenar13, kuan15} or flexible filaments with hard-core interactions~\cite{duman18a, vliegenthart19}, but the collective behavior of active semiflexible filaments with tunable steric interactions has not been explored.

\begin{figure}[tb] \centering
  \includegraphics[width=0.9\textwidth]{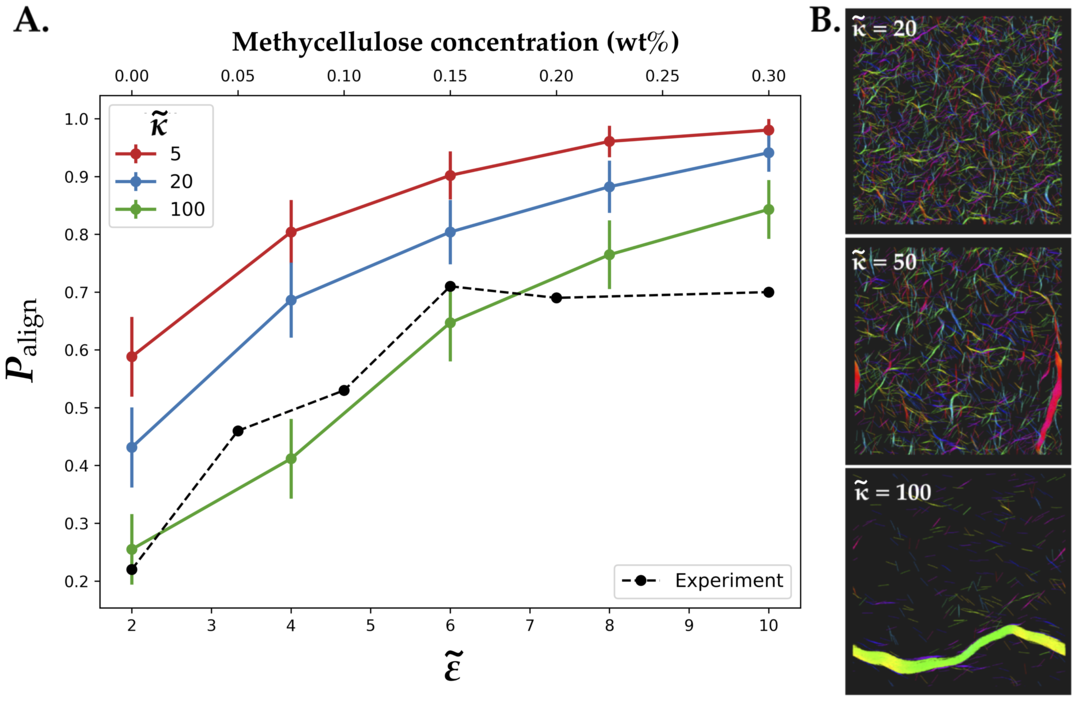}
  \caption{\small A) Plot of the probability for two filaments to align upon collision for $\text{Pe}=10^5$. Increasing repulsion $\tilde{\epsilon}$ increases $P_{\text{align}}$ while increasing rigidity $\tilde{\kappa}$ decreases it. Experimental values taken from Saito et. al~\cite{saito17} with corresponding methylcellulose concentration as the top x-axis. B) Simulation snapshots illustrating how increasing filament rigidity increases collective motion. All images show simulations with $\tilde{\epsilon} = 5$, $\phi=0.1$, $\text{Pe}=10^5$. }
  \label{fig:palign}
\end{figure}

In this paper, we study the non-equilibrium phase behavior of active semiflexible polar filaments that interact with a tunable soft repulsive potential. We find that the dynamic state behavior depends on the strength of the repulsive potential, filament rigidity, and activity. Increasing the repulsion leads to a transition between active isotropic filaments and aligned polar streams, matching the alignment behavior observed in experiments with increasing methycellulose concentration~\cite{inoue15, saito17}. However, we also find that increasing filament rigidity promotes collective motion while simultaneously decreasing the probability that two intersecting filaments align (Fig.~\ref{fig:palign}, ~\cite{supp}). This counterintuitive behavior is explained by an increase in directional persistence tuned by increasing filament rigidity. Therefore, our results suggest that while filament alignment by collisions is important, the degree of filament directional persistence is an additional key physical effect contributing to the emergence of collective motion.

\begin{figure}[tb] \centering
  \includegraphics[width=0.9\textwidth]{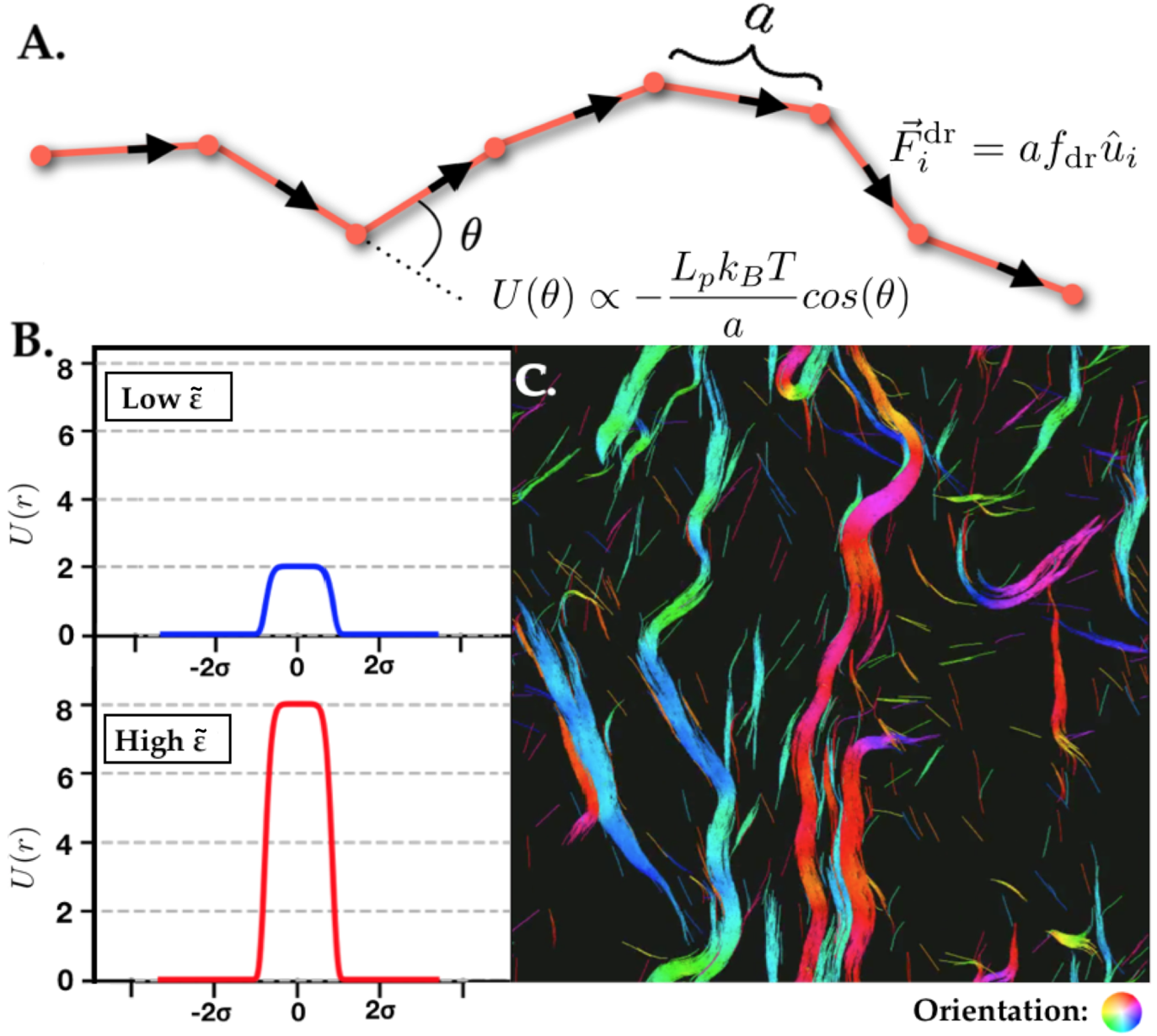}
  \caption{\small Model schematic and simulation snapshot. A) Schematic of self-propelled, semiflexible inextensible, filaments, which contain rigid segments of length $a$ with a bending potential between adjacent segments $U(\theta)$. Each segment experiences a parallel driving force $F_{\text{dr}}$. B) Plot of the repulsive potential $U(r)$ centered on a filament of diameter $\sigma$. C) Snapshot of a simulation with packing fraction $\phi = 0.2$, filament rigidity $\tilde{\kappa} = 100$, and repulsion $\tilde{\epsilon} = 6$.}
  \label{fig:model}
\end{figure}

Our simulation model expands our previous work on self-propelled filaments~\cite{kuan15, gao15, moore20} by adding semiflexibility and tunable repulsion. Our filaments are modeled as inextensible chains of rigid segments, with neighboring segments subject to bending forces to enforce the filament persistence length $L_p$ (Fig.~\ref{fig:model}A). The filament equations of motion are implemented using the constrained Brownian dynamics algorithm of Montesi, Morse, and Pasquali~\cite{montesi05} for a semiflexible chain with anisotropic friction. Each segment of the filament experiences random forces so that its dynamics obey the fluctuation-dissipation theorem for slender filaments at thermal equilibrium~\cite{doi88}. Our choice of filament model has the advantage of being inherently inextensible, and lacks a dependence on large harmonic forces that are typically required to model stiff filaments~\cite{supp}. This allows us to model microtubules, which are inextensible and have high persistence lengths $L_p/L\approx100$--$1000$~\cite{gittes93}, but nevertheless often appear bent in both \emph{in vivo} and \emph{in vitro}, indicating the importance of their flexibility~\cite{odde99, brangwynne06, mcgrath06}.

 The activity of motor proteins is modeled by a polar driving force per unit length $f_\text{dr}$ tangent to each filament segment. This choice reflects experimental observations that filament velocity is constant and independent of methycellulose concentration, even during filament crossing events~\cite{liu11, kim18}, suggesting that stochastic effects due to the motors (e.g., binding and unbinding, variation in motor stepping) occur at short enough time scales so as to not significantly alter large-scale behavior. 

Interactions between filaments are repulsive but soft, defined by the generalized exponential potential (GEM-8), $U(r) = \epsilon e^{-(r/\sigma)^8} $, with cutoff $U(r>\sqrt{2}\sigma) = 0$. Here $r$ is the minimum distance between neighboring segments and $\sigma$ the diameter of a filament (Fig.~\ref{fig:model}B). The maximum potential value $\epsilon$ represents the energy required for two segments to overlap. This potential is steep near the edge of a filament. Interactions occur between each filament segment, except for nearest-neighbor segments of the same filament. While filaments experience local drag, we neglect long-range hydrodynamic interactions because previous experiments found these forces to be negligible~\cite{saito17}.

Filaments are inserted at a packing fraction $\phi = A_\text{fil}/A_\text{sys}$, where $A_{\text{sys}}$ is the area of the simulation box and $A_{\text{fil}}=N(L\sigma + \pi\sigma^2)$ is the total area occupied by $N$ spherocylindrical filaments of length $L$ and diameter $\sigma$. The characteristic timescale is the time for a filament to move the distance of its contour length, $\tau_A = L/v$, with the velocity depending on the total driving force and the coefficient of friction acting parallel to the filament, $v = \xi_{\parallel}^{-1} F_{\text{dr}}$. The filament driving force per unit length is set by the P\'eclet number, which is the ratio of active and diffusive transport time scales, $\text{Pe} = \tau_D/\tau_A = f_{\text{dr}} L^2/k_B T$. We explore the range $\text{Pe} = 10^4$--$10^5$, based on calculations of active forces in experiments~\cite{liu11, supp}. Simulations were run for $10^3-10^4 \tau_A$.

Our results depend on six dimensionless parameters: rigidity $\tilde{\kappa}=L_p/L$, interaction energy $\tilde{\epsilon} = \epsilon/\epsilon_{\text{dr}}$, packing fraction $\phi = A_{\text{fil}}/A_{\text{sys}}$, aspect ratio $l=L/\sigma=60$, system size $l_{sys}=L_{sys}/L=20$, and P\'eclet number $\text{Pe}$. The interaction energy has been rescaled by $\epsilon_{\text{dr}}$, which is the energy required for the potential to exert a maximum force equal to the driving force of a particle with length $\sigma$. Our simulation varies $\tilde{\kappa}$ from $5$--$1000$; previous computational work on active semiflexible filaments examined the range $\tilde{\kappa} < 20$~\cite{duman18a, vliegenthart19}, however this is not well-suited for the study of microtubules, which can have $\tilde{\kappa} \approx 100$--$1000$. We varied $\tilde{\epsilon}$ from 1--20, corresponding to filament pair-alignment behavior estimated from previous gliding assay experiments~\cite{inoue15, saito17, kim18} (Fig.~\ref{fig:palign}). The work here focuses on the low-density regime of $\phi = 0.05$, $0.1$, but we also explored higher filament densities ($\phi = 0.2$, $0.4$, and $0.8$) at $\text{Pe} = 10^5$~\cite{supp}. Although the filament aspect ratio is fixed, varying the filament rigidity is analogous to varying the length of a filament with a fixed persistence length.

\begin{figure*}[t] \centering
  \includegraphics[width=0.9\textwidth]{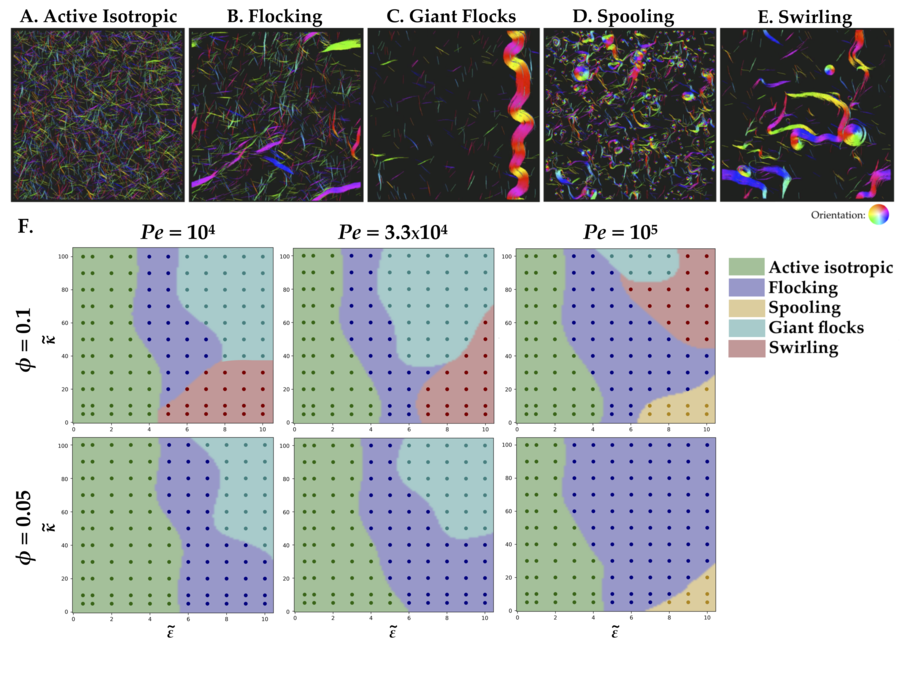}
  \caption{\small Simulation snapshots and phase diagram. A) Active isotropic ($\phi=0.2$, $\tilde{\epsilon}=1.5$, $\tilde{\kappa}=50$). B) Flocking ($\phi=0.2$, $\tilde{\epsilon}=2$, $\tilde{\kappa}=100$). C) Polar band ($\phi=0.2$, $\tilde{\epsilon}=5$, $\tilde{\kappa}=100$). D) Spooling ($\phi=0.2$, $\tilde{\epsilon}=10$, $\tilde{\kappa}=20$). E) Swirling ($\phi=0.2$, $\tilde{\epsilon}=10$, $\tilde{\kappa}=100$). F) Phase diagram of self-propelled semiflexible filaments with varying rigidity $\tilde{\kappa}$ and repulsion $\tilde{\epsilon}$ for P\'eclet number $\text{Pe} = 10^4$--$10^5$, and packing fraction $\phi = 0.05$--$0.1$. Points represent simulations; background colors are for visualization of clusters. Separation of phases was achieved through high-dimensional clustering of the simulation order parameters~\cite{supp}.}
  \label{fig:examples}
\end{figure*}

Our simulations generated five primary phases: active isotropic, flocking, giant flocking, swirling, and spooling (Fig.~\ref{fig:examples}). To quantify the dynamical phases, we used six global order parameters: the polar order $P$, nematic order $Q$, average contact number $c$, average local polar order $p$, average spiral number $s$, and number fluctuations $\Delta N$~\cite{supp}. We also quantified the collective dynamics of the system by characterizing the flocking behavior in terms of the number of flocking filaments $N_\text{F}/N$, and frequencies that filaments joined or left a flocking state, respectively $f_\text{NF--F}$ and $f_\text{F--NF}$~\cite{supp}. Using a high-dimensional clustering algorithm, we identified 5 clusters in this nine-dimensional order parameter space corresponding to the different phases observed for our range of parameters~\cite{supp}.

In the active isotropic phase, filaments cross each other in all directions, resembling filament gliding experiments that do not exhibit collective motion (Fig.~\ref{fig:examples}A). The flocking phase is characterized by the coexistence of multiple polar domains of aligned filaments (Fig.~\ref{fig:examples}B). Flocks are dynamic, with filaments continuously joining and leaving the flocking state. The giant flocking phase occurs when all flocks in the system coalesce into a single dominating flock, and exhibits long-range order. At higher densities, giant flocks can span the system length (Fig.~\ref{fig:examples}C), and can be either one band ($\phi=0.1$--$0.2$) or many counter-propagating bands ($\phi=0.4$--$0.8$). In the spooling phase, filaments are flexible enough to self-interact, and many filaments form spirals (Fig.~\ref{fig:examples}D). The swirling phase contains large, swarming flocks that collide, self-interact, and form transient vortices (Fig.~\ref{fig:examples}E).

Collective motion emerges with increasing repulsion or rigidity. When both filament repulsion and rigidity are small, we find the active isotropic phase at all densities and P\'eclet numbers (Fig.~\ref{fig:examples}F). Self-organization and collective motion can occur by increasing either repulsion or stiffness; increasing $\tilde{\epsilon}$ tends to increase collective behavior, while increasing $\tilde{\kappa}$ or $\phi$ tends to increase long-range order. Previous experimental work has found that the transition from an active isotropic phase to polar streams was driven by an increase in alignment events between pairs of colliding filaments~\cite{sumino12, inoue15, saito17, kim18}. Accordingly, we measured the correlation between the emergence of collective behavior and the probability of filament alignment, $P_{\text{align}}$~\cite{supp}. We find that $P_{\text{align}}$ increases monotonically with increasing repulsion, in agreement with trends in experiments for increasing methylcellulose concentrations~\cite{inoue15, saito17, kim18} (Fig.~\ref{fig:palign}). In contrast, $P_{\text{align}}$ decreases monotonically with increasing stiffness~\cite{supp}. Remarkably, we found regions of phase space where increasing stiffness drives the emergence of collective motion, and regions where increasing repulsion inhibits long-range organization and collective behavior.

The negative correlation between $P_{\text{align}}$ and filament rigidity can be understood by considering the additional energy required to apply a torque on a stiff filament. Flexible filaments need only bend very marginally at the leading end to align during a collision. Stiff filaments cannot bend as easily and thus require additional torque to align. Higher torques between colliding filaments would increase overall alignment in the case of hard-core repulsive interactions~\cite{baskaran08}. However, with a finite repulsivity, the higher energy cost leads to a reduction in the probability of alignment.

\begin{figure}[tb] \centering
  \includegraphics[width=0.9\textwidth]{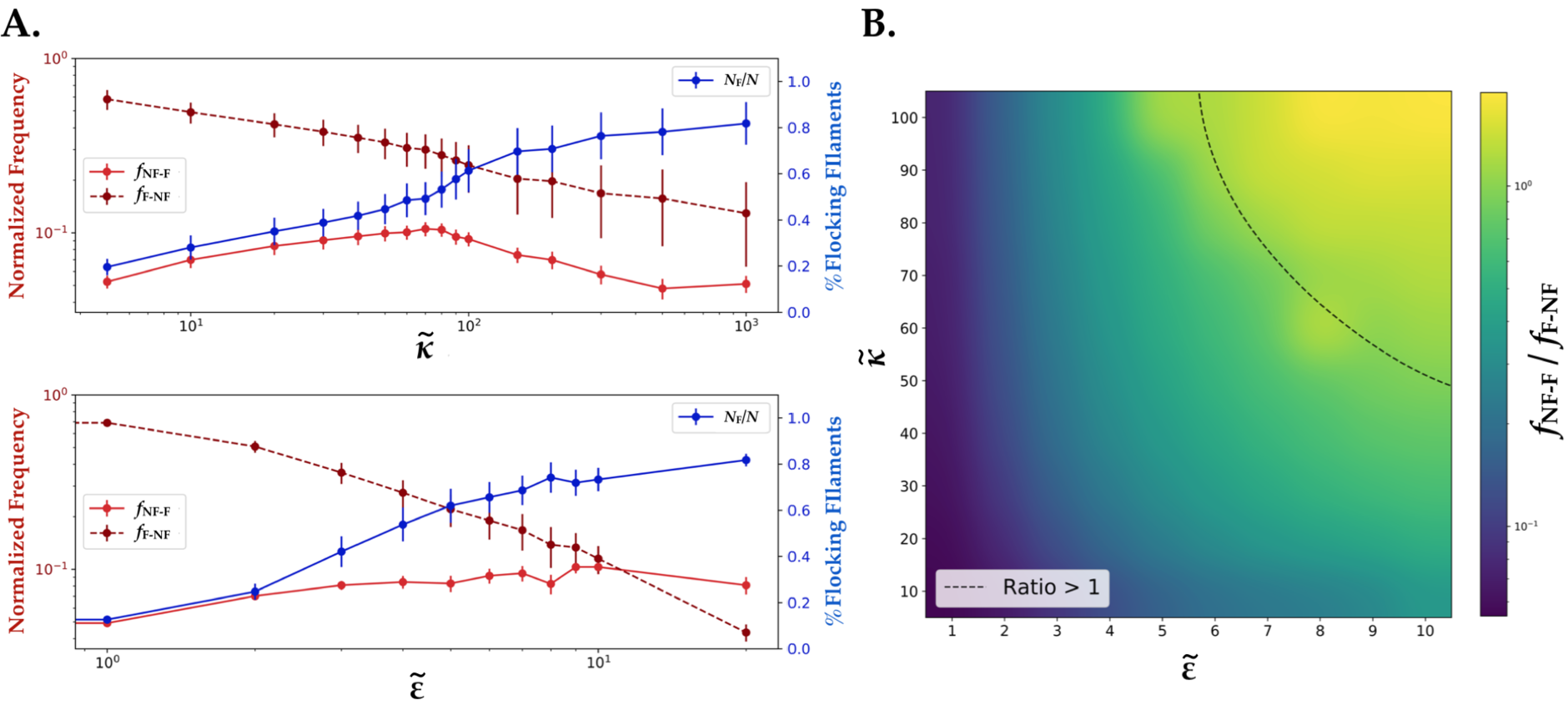}
  \caption{\small A) Switching frequency between the flocking (F) and not-flocking (NF) states plotted as a function of filament rigidity (top) and soft repulsion strength (bottom) averaged over all simulations with $\phi=0.1$, $\text{Pe}=10^5$, and time-averaged for the final $10\%$ of the simulation. The frequency is normalized by the population of filaments available in the initial state, and plotted alongside the percentage of filaments that are flocking in the system. The frequencies have units $\tau_A^{-1}$~\cite{supp}. B) Plot of the ratio of the flocking state switching frequency ratio $f_\text{NF--F}/f_\text{F--NF}$ with respect to $\tilde{\kappa}$ and $\tilde{\epsilon}$ for $\phi=0.1$, $\text{Pe}=10^5$. The region where the frequency for joining a flock is greater than the departure frequency is indicated by the dashed line.}
  \label{fig:flock}
\end{figure}

The counterintuitive result that increasing stiffness lowers collisional alignment but increases collective motion reflects a trade-off between filament alignment and the directional persistence of filament trajectories. We quantified the directional persistence of filaments by measuring the autocorrelation of filament orientation $u$, $C(t) = \langle u(0) \cdot u(t) \rangle$, at different rigidities and P\'eclet numbers in the absence of interactions~\cite{supp}. Since driven filaments are readily deflected from their trajectories by small deviations of the leading filament tip, rigid filaments that resist deflection exhibit longer orientation correlation times. Flocking filaments that move ballistically have longer-lived alignment and thus a higher flocking lifetime, permitting other filaments to join. Although the frequency of alignment may be lower for stiff filaments, the longer duration of alignment compensates for the low alignment probability. In contrast, flexible filaments have a higher rate of alignment, leading to the rapid creation of polar clusters (Fig.~\ref{fig:flock}), but they are easily deflected away, resulting in short-lived flocks. Thus, flexible filaments require a larger $\tilde{\epsilon}$ to cross the active isotropic--flocking boundary. Our results suggest that filament alignment and directional persistence are both important for the emergence of collective motion and stable flocks.

The flocking phase consists of multiple polar domains of aligned filaments, with individual flocks characterized by high local filament density and local polar order $p$. Individual flocking filaments are identified by the criterion $p_i > 0.5$, with filaments that are located in the flock interior exhibiting high contact number $c_i$ as well. The number of flocking filaments increases with filament stiffness and repulsivity. Filaments with lower $\tilde{\kappa}$ have faster switching between flocking and non-flocking states compared to stiff filaments, yet the saturation of the number of flocking filaments occurs at high $\tilde{\kappa}$ and slow switching frequencies (Fig.~\ref{fig:flock}).

With increasing stiffness, flocks that form in a transient flocking phase can coalesce into a single dominating flock, characteristic of the giant flock phase. In the giant flock phase, the number of flocking filaments saturates, depriving the remaining system of the necessary filament density to form additional stable flocks. Giant flocks in this phase are characteristically long and narrow, allowing them to efficiently intersect with and capture non-flocking filaments. The rate of switching between flocking and non-flocking states becomes low in the giant flock phase, due to the large number of filaments that are kinetically trapped in the flock interior~\cite{supp}. Giant flocks are more stable at lower driving, consistent with previous reports that high activity can inhibit collective behavior of filaments with flexibility~\cite{duman18a}. Due to finite-size effects of the simulation, the continued growth of the giant flock at higher densities ($\phi \geq 0.1$) will often lead it to span the length of the system, $L_\text{flock} > 20 L$, ending with the formation of a persistent polar band. At very high densities ($\phi \geq 0.4$), multiple independent bands can form simultaneously, resulting in coexisting nematic bands. We note that high filament density paired with high $\tilde{\kappa}$ results in a stable giant flock phase even at low $\tilde{\epsilon}$. The high local filament densities serve to restrict the rotational degrees of freedom of gliding filaments, further increasing directional persistence and the stability of polar bands.

Our phase diagram shows a limited region of stability for long-range collective motion in the giant flocking phase. This occurs because although increasing $\tilde{\kappa}$ promotes long-range order, increasing $\tilde{\epsilon}$ too high can break it. At high $\tilde{\kappa}$ and high $\tilde{\epsilon}$, collisions cause large deformations of flocking and banding filaments, leading to buckling. In the simulation shown in Fig.~\ref{fig:examples}C, the filaments repeatedly form a polar band, which buckles, shears, falls apart, and eventually reforms. Similar dynamic phases have been observed in systems of self-propelled rods~\cite{ginelli10} and semiflexible filaments driven by motors~\cite{vliegenthart19}. Therefore, only intermediate values of $\tilde{\epsilon}$ facilitate persistent long-range order as giant flocks.

\begin{figure}[tb] \centering
  \includegraphics[width=0.9\textwidth]{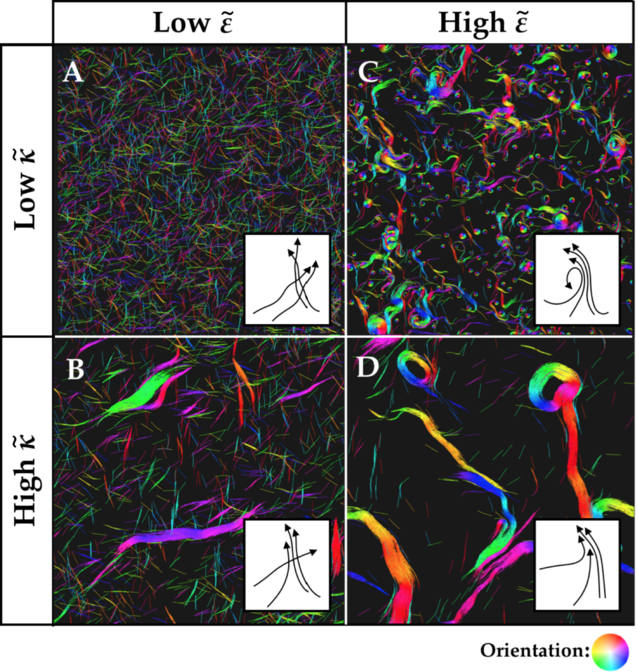}
  \caption{\small Diagram depicting the phase behavior of active filaments with varying rigidity $\tilde{\kappa} = 20, 100$ and repulsion $\tilde{\epsilon}=2, 10$ at $\phi = 0.2$ and $\text{Pe}=10^5$. Increasing repulsion decreases the probability of filament crossing. Increasing filament rigidity increases polar order of flocks and increases resistance to filament bending in collisions. }
  \label{fig:behavior}
\end{figure}

While increasing $\tilde{\epsilon}$ typically increases alignment events, promoting collective motion, for flexible filaments (low $\tilde{\kappa}$), filaments can bend and self-interact via collisions to form spirals (Fig.~\ref{fig:behavior}C). Spirals are seen at all P\'eclet numbers, but are increasingly stable with higher driving. Stable single-filament spirals can persist until a collision with another filament or flock deforms the filament enough to release it. At high driving, the system can enter the spooling phase, wherein a majority of filaments become kinetically trapped as spirals. The spooling phase resembles the frozen, active steady states found in previous experimental work~\cite{schaller11}. The principle mechanism of spool formation has been of significant interest~\cite{liu11, luria11, t.lam14, isele-holder15, vandelinder16a, duman18a, mokhtari19}, with suggestions ranging from defects in the motor lattice to thermal activation.  Previous experiments with microtubules and dynein have observed stable filament vortices attributed to filament curvature induced by motors~\cite{sumino12}, which are absent in our simulations. The spooling phase demonstrated here is governed by steric self-interactions and collisions, as found in previous modeling work~\cite{isele-holder15, duman18a}. Their overwhelming appearance at high driving may be due to a rescaling of the effective filament rigidity at high activity, which has been reported in other recent work~\cite{isele-holder15, singh18, gupta19, peterson20, supp}, leading to an apparent softening of the filament that may contribute to a buckling instability~\cite{isele-holder16}. Therefore, while spool formation in experiments may have contributions from defects in the protein lattice, pinning of filaments by dead motors, or intrinsic curvature, our results indicate that these mechanisms are not necessary for spool formation.

When repulsion and stiffness are both high, even transient polar bands can no longer form, and the system enters the swirling phase. High-energy collisions due to large bending and repulsive forces cause large deformation of flocks, leading to shorter end-to-end flock length and inhibiting the long-range order that is present at lower interaction energy. Flocks in the swirling phase can have much shorter aspect ratios compared to giant flocks. The sharp bending of flocks can result in flock self-interaction, which may form a large, transient vortex of filaments (Fig.~\ref{fig:behavior}D).

Our simulations do not display symmetry breaking of the system chirality as observed in some previous work~\cite{schaller11, sumino12, kim18}, due to zero preferred filament curvature in our model. When an intrinsic curvature is added with simulation parameters that otherwise form stable polar bands, we observe a rotation in the polar order vector~\cite{supp}, similar to previous simulations of gliding filaments with intrinsic curvature~\cite{kim18}.

We have examined of the role of flexibility and repulsivity in the collective behavior of active polar filaments. The phase diagram presented here makes predictions that could guide future experiments seeking to observe collective behavior in systems of active semiflexible polymers. These systems can exhibit a wide variety of active, dynamic states, with transitions that are controllable by tuning repulsive interactions and filament rigidity. The ability to control the transition between these states may have applications for drugs targeting cortical cytoskeletal filaments or nanodevices that use cytoskeletal filaments as molecular shuttles~\cite{nitta06, hess11}.

We would like to thank Michael Stefferson for helpful discussions. This work was funded by the Soft Materials Research Center under NSF MRSEC Grant No. DMR-1420736, and the National Science Foundation under NSF GRFP Award No. DGE-1144083 and NSF Grant No. DMR-1725065. This work utilized the RMACC Summit supercomputer, which is supported by the National Science Foundation (awards ACI-1532235 and ACI-1532236), the University of Colorado Boulder, and Colorado State University. The Summit supercomputer is a joint effort of the University of Colorado Boulder and Colorado State University.



\bibliographystyle{apsrev4-1}
\bibliography{cmdsf}
\end{document}



\title{\bf{Supplemental Material: Collective motion of driven semiflexible filaments tuned by soft repulsion and stiffness}} 
\author{Jeffrey M. Moore}
\author{Tyler N. Thompson}
\author{Matthew A. Glaser} 


\affiliation{Department of Physics, University of Colorado, Boulder, CO 80309}

\author{Meredith D. Betterton}
\affiliation{Department of Physics, University of Colorado, Boulder, CO 80309}
\affiliation{Department of Molecular, Cellular, and Developmental Biology, University of Colorado, Boulder, CO 80309}

\date{\today}

\maketitle

\section{Simulation model}
Our semiflexible filaments are modeled as discretized wormlike chains~\cite{kratky49} with rigid, inextensible segments. We adopt the algorithm for constrained Brownian dynamics of bead-rod wormlike chains with anisotropic friction by Montesi et. al~\cite{montesi05}. The details that follow are an overview of the algorithm introduced in their paper, with the addition of interactions and self-propulsion forces.

Filaments are represented by $N$ sites and $N-1$ segments, with fixed segment length $a$, contour length $L=(N-1)a$, and anisotropic friction, $\zeta_{\bot} = 2\zeta_{\parallel}$. The position of each site $\mathbf{r}_i$ is updated using a midstep algorithm

\begin{equation}\label{posupdate}
  \begin{aligned}
    \mathbf{r}_i^{(1/2)} &= \mathbf{r}_i^{(0)} + \frac{\Delta t}{2}\mathbf{v}_i^{(0)} , \\ 
    \mathbf{r}_i^{(1)} &= \mathbf{r}_i^{(0)} + \Delta t\ \mathbf{v}_i^{(1/2)} ,
  \end{aligned}
\end{equation}
where $\Delta t$ is the time step, $\mathbf{v}_i^{(0)}$ is the initial velocity of site $i$ at initial position $\mathbf{r}_i^{(0)}$, and $\mathbf{v}_i^{(1/2)}$ is the velocity of site $i$ recalculated at the midstep position $\mathbf{r}_i^{(1/2)}$, but using the same stochastic forces as in $\mathbf{v}_i^{(0)}$. Each site $i$ is assigned an orientation, corresponding to the orientation of the segment attaching it to site $i+1$,

\begin{equation}
  \mathbf{u}_i = \frac{\mathbf{r}_{i+1} - \mathbf{r}_i} { | \mathbf{r}_{i+1} - \mathbf{r}_i | } = \frac{1}{a} (\mathbf{r}_{i+1} - \mathbf{r}_i).
\end{equation}
The orientation of the last site of the filament is set equal to that of its only neighboring segment, so that $\mathbf{u}_N = \mathbf{u}_{N-1}$.

The velocity of each site is
\begin{equation}
  \mathbf{v}_i = \mathbf{H}_{ij} \cdot \mathbf{F}^{\text{tot}}_j ,
\end{equation}
where $\mathbf{H}_{ij}$ is an anisotropic mobility tensor, and is the mobility of site $i$ in response to a force on site $j$. The total force on site $i$ is the sum

\begin{equation}
  \mathbf{F}^{\text{tot}}_i = \mathbf{F}_i^{\text{bend}} + \mathbf{F}_i^{\text{metric}} + \mathbf{F}_i^{\text{tension}} + \mathbf{F}_i^{\text{ext}} + \bm{\eta}_i.
\end{equation}
The deterministic forces include filament bending forces, metric forces, tension forces, and external forces from interactions. $\bm{\eta}_i$ are the random forces contributing to the Brownian motion of the filament, and are geometrically-projected such that the forces due not break the constraints due to the fixed segment length, and are described in detail by Montesi et al.~\cite{montesi05}.

The mobility tensor can be written as an inverse site friction tensor

\begin{equation}
  \begin{aligned}
    \mathbf{H}_{ij} &= \delta_{ij}\bm{\zeta}^{-1}_j , \\
    \bm{\zeta}^{-1}_i &= \frac{1}{\zeta_{\parallel}^i} \tilde{\mathbf{u}}_i \otimes \tilde{\mathbf{u}}_i + \frac{1}{\zeta_{\bot}^i} \big(\mathbf{I} - \tilde{\mathbf{u}}_i \otimes \tilde{\mathbf{u}}_i \big) .
  \end{aligned}
\end{equation}
where $\tilde{\mathbf{u}}_i$ is a vector tangent to site $i$, the $\otimes$ symbol denotes the outer product, and the parallel and perpendicular friction coefficients corresponding to site $i$ are $\zeta_{\parallel}^{i}$ and $\zeta_{\bot}^i$ respectively. The tangent vector is the average of the orientations $\mathbf{u}_i$ of its neighboring segments,

\begin{equation}\label{utan}
    \tilde{\mathbf{u}}_i = \frac{(\mathbf{u}_i + \mathbf{u}_{i-1})}{  |\mathbf{u}_i + \mathbf{u}_{i-1} | }
\end{equation}
for $2 \leq i \leq N$, and $\tilde{\mathbf{u}}_1 = \mathbf{u}_1$, $\tilde{\mathbf{u}}_N = \mathbf{u}_{N-1}$ at the chain ends. The position update routine in Eqn.~\ref{posupdate} can be rewritten in terms of the inverse site friction tensor as

\begin{align}
  \mathbf{r}_i^{(1/2)} &= \mathbf{r}_i^{(0)} + \frac{\Delta t}{2}\bm\zeta_i^{-1,(0)}\cdot\mathbf{F}^{\text{tot},(0)}_i , \label{midstep} \\ 
  \mathbf{r}_i^{(1)} &= \mathbf{r}_i^{(0)} + \Delta t\ \bm\zeta_i^{-1,(1/2)}\cdot\mathbf{F}^{\text{tot},(1/2)}_i, \label{fullstep}
\end{align}

The diffusivity of the wormlike chain is $D = k_B T/\zeta = k_B T/N\zeta^i$, where $\zeta^i$ is the local friction due to site $i$, which depends on the filament aspect ratio $L/d$ where $d$ is the diameter of the chain. In the regime of rigid, infinitely thin rods, the coefficient of friction is given by~\cite{doi88},

\begin{equation}
  \lim_{L/d \to \infty} \zeta_{\bot} = 4 \pi \eta_s L \epsilon.
\end{equation}
where $\epsilon = 1/\ln{(L/d)}$. In the finite aspect ratio case, this friction coefficient is multiplied by a geometric factor
\begin{equation}
  f(\epsilon) = \frac{1+0.64\epsilon}{1-1.15\epsilon}+1.659\epsilon^2.
\end{equation}
Therefore, each site experiences a local friction given by
\begin{equation}
  \zeta_{\bot}^{\text{i}} = 4 \pi \eta_s a \epsilon f(\epsilon).
\end{equation}

The bending energy of a discrete wormlike chain for $N \gg 1$ is approximated by

\begin{equation}\label{ubend}
  U_{\text{bend}} = - \frac{\kappa}{a}\sum_{k=2}^{N-1}\mathbf{u}_k\cdot\mathbf{u}_{k-1},
\end{equation}
where $\kappa$ is the bending rigidity, which is related to the persistence length $L_p$ of the wormlike chain as $\kappa/k_B T = L_p$. Note that we are adopting the convention that the previous equation is true in all dimensions $d$ of wormlike chains, unlike the convention adopted by Landau and Lifshitz where $\kappa/k_BT = (d-1)L_p/2$~\cite{landau86}. This results in a Kuhn length that depends on dimensionality, $b = (d-1) L_p$, which is critical when analyzing the correlations of the wormlike chain for $d=2$, as discussed later.

The bending force is $\mathbf{F}^{\text{bend}}_i=-\partial U_{\text{bend}}/\partial \mathbf{r}_i$. The implementation of the bending forces coincides with metric forces, which come from a metric pseudo-potential

\begin{equation}
  U_{\text{metric}} = \frac{k_B T}{2} \ln{(\det \hat{\mathbf{G}})},
\end{equation}
and $\hat{\mathbf{G}}$ is the metric tensor, which is an $(N-1)\times(N-1)$ tridiagonal matrix. $\hat{\mathbf{G}}$ has diagonal terms $d_i = 2$ and the off-diagonal terms depend on the cosine of the angle between neighboring segments, $c_i = -\mathbf{u}_i\cdot\mathbf{u}_{i-1}$. The metric pseudo-potential is necessary for the filament conformation to have the expected statistical behavior in the flexible limit, $L_p \ll L$.

It was shown in the work of Pasquali et al.~\cite{pasquali02} that the bending forces and metric forces could be calculated together as one force term,
\begin{equation}\label{eqn:fbendfmetric}
  \mathbf{F}_i^{\text{bend}}+\mathbf{F}_i^{\text{metric}} = \frac{1}{a}\sum_{k=2}^{N-1}\kappa_k^{\text{eff}}\frac{\partial (\mathbf{u}_k\cdot\mathbf{u}_{k-1})}{\partial \mathbf{r}_i},
\end{equation}
where $\kappa^{\text{eff}}$ replaces the true bending rigidity $\kappa$ as an effective rigidity with a conformational dependence,
\begin{equation}\label{keff}
  \kappa_i^{\text{eff}} = \kappa + k_B T a \hat{G}^{-1}_{i-1,i}.
\end{equation}
The derivative in Eqn.~\ref{eqn:fbendfmetric} can be evaluated from
\begin{equation}
  \frac{\partial \mathbf{u}_k}{\partial \mathbf{r}_i} = \frac{1}{a} (\delta_{i,k+1} - \delta_{i,k})(\mathbf{I}-\mathbf{u}_k\otimes\mathbf{u}_k), 
\end{equation}
so the two forces can be written
\begin{equation}\label{eqn:fbendfmetricx}
  \mathbf{F}_i^{\text{bend}}+\mathbf{F}_i^{\text{metric}} = \frac{1}{a^2}\sum_{k=2}^{N-1}\kappa_k^{\text{eff}}
  \Big((\delta_{i,k+1} - \delta_{i,k})(\mathbf{I}-\mathbf{u}_k\otimes\mathbf{u}_k)\mathbf{u}_{k-1} + (\delta_{i,k} - \delta_{i,k-1})(\mathbf{I}-\mathbf{u}_{k-1}\otimes\mathbf{u}_{k-1})\mathbf{u}_{k}\Big).
\end{equation}

For an inextensible wormlike chain, the positions of $N$ sites must satisfy $N-1$ constraints $C_\mu$ where
\begin{equation}
  C_\mu = | \mathbf{r}_{\mu+1} - \mathbf{r}_{\mu} | = a,
\end{equation}
for $\mu = 1,\dots,N-1$. Differentiating the constraints with respect to the site positions $\mathbf{r}_i$ yields a vector
\begin{equation}
  \mathbf{n}_{i\mu} = \mathbf{u}_{\mu} (\delta_{i,\mu+1} - \delta_{i,\mu}).
\end{equation}
The geometrically projected random forces $\bm{\eta}_i$ must satisfy the property
\begin{equation}
  \bm{\eta}_i \cdot \mathbf{n}_{i\mu} = 0,
\end{equation}
so that the $3N$ dimensional vector of random forces $\bm{\eta}_i$ is locally tangent to the $3N - (N-1) = 2N+1$ dimensional hypersurface to which the system is confined. 

The geometrically projected random forces $\bm{\eta}$ are calculated from
\begin{equation}\label{geoprojrand}
  \begin{aligned}
  \bm{\eta}_i &= \bm{\eta}'_i - \mathbf{n}_{i\mu}\hat{\eta}_{\mu}, \\
  &= \bm{\eta}'_i + \hat{\eta}_{i}\mathbf{u}_i - \hat{\eta}_{i-1}\mathbf{u}_{i-1}, 
  \end{aligned}
\end{equation}
where $\bm{\eta}'_i$ are the unprojected random forces and $\hat{\eta}_{\mu}$ is the component of the $3N$ dimensional unprojected random force vector along direction $\mathbf{n}_{i\mu}$. The unprojected random forces at each timestep are
\begin{equation} \label{unprojrand}
  \begin{aligned}
    \bm{\eta}'_i &= \sqrt{24 k_B T/\Delta t} \bm{\zeta}^{1/2}_i \cdot \bm{\xi}_i \\
    &=  \sqrt{24 k_B T/\Delta t} \Big( \zeta^{1/2}_{\bot}\bm{\xi}_i + (\zeta^{1/2}_{\parallel} - \zeta^{1/2}_{\bot}) \tilde{\mathbf{u}}_i \otimes \tilde{\mathbf{u}}_i \cdot \bm{\xi}_i \Big) ,
  \end{aligned}
\end{equation}
where $\bm\xi_i$ is a spatial vector whose elements are uniformly distributed random numbers between $-0.5,0.5$~\cite{montesi05, grassia95}.

To calculate $\hat\eta_i$, we solve the set of $N-1$ linear equations

\begin{equation}\label{hardcomp}
  \sum_{\nu=1}^{N-1}\hat{G}_{\mu\nu}\hat{\eta}_{\nu} = (\bm\eta'_{\mu+1}-\bm\eta'_{\mu})\cdot\mathbf{u}_{\mu} = p_{\mu},
\end{equation}
where $\hat{\mathbf{G}}$ is the metric tensor. In matrix notation, this is equivalent to solving $\hat{\mathbf{G}}\hat{\bm{\eta}}=\mathbf{p}$ for $\hat{\bm{\eta}}$, and can be solved in $\mathcal{O}(N)$ steps using LU decomposition for tridiagonal matrices. Once the system is solved for $\hat{\eta}_i$, we use Eqn.~\ref{geoprojrand} to solve for the geometrically projected random forces. While the deterministic forces are recalculated and applied at each half step of the simulation, the geometrically projected random forces are only calculated once per full simulation step at the initial site positions $\mathbf{r}_i^{(0)}$, and applied at each half step of the algorithm.

To calculate the tension $\mathcal{T}_i$, we require that the system constraints are constant, i.e. $\dot{C}_{\mu} = 0$. This is equivalent to solving the system of linear equations 

\begin{equation}\label{htauq}
  \sum_{\nu=1}^{N-1}\hat{H}_{\mu\nu} \mathcal{T}_{\nu} = \mathbf{u}_{\mu}\cdot (\bm{\zeta}^{-1}_{\mu+1}\cdot\mathbf{F}^{\text{uc}}_{\mu+1} - \bm{\zeta}^{-1}_{\mu}\cdot\mathbf{F}^{\text{uc}}_{\mu}) = q_{\mu},
\end{equation}

\noindent where $\mu = 1,\dots,N-1$. In matrix notation, this can be written as $\hat{\mathbf{H}}\bm{\mathcal{T}} = \mathbf{q}$, where $\hat{\mathbf{H}}$ is another tridiagonal $(N-1)\times(N-1)$ matrix

\begin{equation}
  \hat{H}_{\mu\nu} = \sum_{i=1}^{N} \mathbf{n}_{i\mu}\cdot\bm{\zeta}^{-1}_i\cdot\mathbf{n}_{i\nu}
\end{equation}
or
\begin{equation}
  \hat{\mathbf{H}}=
  \begin{bmatrix}
    b_1 & a_2 & 0 & \dots & 0 & 0 \\
    a_2 & b_2 & a_3 & 0 & \dots  & 0 \\
    0 & a_3 & b_3 & a_4 & 0 & \dots \\
    \hdotsfor{6} \\
    \dots & 0 & a_{N-3} & b_{N-3} & a_{N-2} & 0 \\
    0 & \dots & 0 & a_{N-2} & b_{N-2} & a_{N-1} \\
    0 & 0 & \dots & 0 & a_{N-1} & b_{N-1} 
  \end{bmatrix},
\end{equation}
with diagonal and off-diagonal elements
\begin{equation}
  \begin{aligned}
    b_{\mu} &= \frac{2}{\zeta_{\bot}} + \Big(\frac{1}{\zeta_{\parallel}}-\frac{1}{\zeta_{\bot}}\Big) \Big( (\tilde{\mathbf{u}}_{\mu} \cdot \mathbf{u}_{\mu})^2 + (\tilde{\mathbf{u}}_{\mu+1}\cdot\mathbf{u}_{\mu})^2\Big), \\
    a_{\mu} &= -\frac{1}{\zeta_{\bot}}\mathbf{u}_{\mu+1}\cdot\mathbf{u}_{\mu} - \Big(\frac{1}{\zeta_{\parallel}}-\frac{1}{\zeta_{\bot}}\Big) \Big( (\tilde{\mathbf{u}}_{\mu}\cdot\mathbf{u}_{\mu+1})(\tilde{\mathbf{u}}_{\mu}\cdot\mathbf{u}_{\mu})\Big) .
  \end{aligned}
\end{equation}
and can be solved in a similar manner as Eqn.~\ref{hardcomp}. The tension forces are then
\begin{equation}\label{ftension}
  \mathbf{F}_i^{\text{tension}} = \mathcal{T}_i\mathbf{u}_i - \mathcal{T}_{i-1}\mathbf{u}_{i-1}.
\end{equation}

External forces $\mathbf{F}^{\text{ext}}_i$ include forces from filament-filament interactions and self-propulsion forces from the driving of molecular motors. The filament-filament interaction forces are calculated from the derivative of the general exponential model potential (GEM-8)
\begin{equation}
  U(r) = 
  \begin{cases}
    \epsilon e^{-(r/\sigma)^8} &\quad\text{if } r<\sqrt{2}\sigma, \\
    0 &\quad\text{otherwise.}
  \end{cases}
\end{equation}
where $r$ is the minimum distance between neighboring filament segments and $\sigma$ is the unit length used in the simulation, defined to be the diameter of a filament. 

Forces from molecular motors are modeled in our simulations as a uniform force density $f_{\text{dr}}$ that is directed along the local filament segment orientations,
\begin{equation}
  \mathbf{F}_{\text{dr}} = f_{\text{dr}}\mathbf{u}_i.
\end{equation}
The assumptions of this model are that lattice defects are negligible for observing collective behavior of gliding filaments, and that motor binding and unbinding events occur at fast enough rates such that their behavior need not be explicitly included in the model. These assumptions are guided by experimental observations that filament velocities are constant in gliding assays, despite the presence of filament crossing events that certainly require a large number of unbinding and binding events~\cite{liu11}.

\section{Model implementation}

Simulation software for the filament model is written in C++ and is publicly available online~\cite{moore20}. The simulations were run on the Summit computing cluster~\cite{anderson17} and parallelized using OpenMP. The simulations presented in this work required approximately $10^6$ CPU hours of computation, as a conservative estimate, plus additional resources for post-processing and analysis.

\section{Model validation}

\begin{figure*}[htb] \centering
  \includegraphics[width=\textwidth]{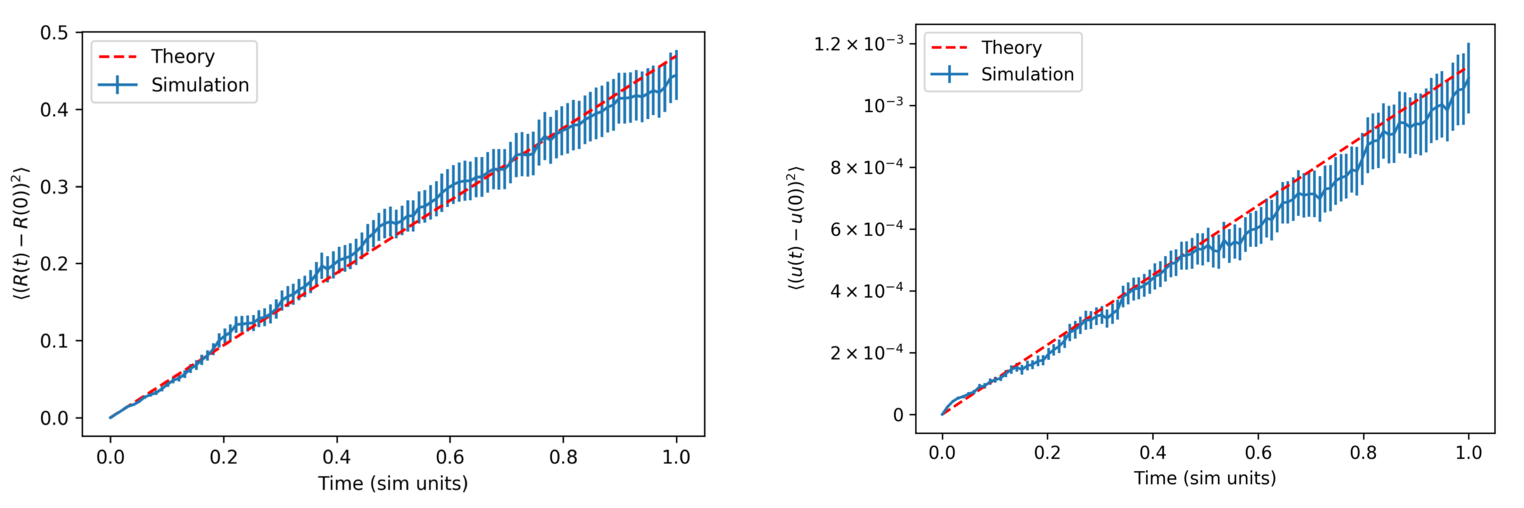}
  \caption{\small Left, simulation and theoretical comparison of the mean-squared displacement (MSD) averaged over $100$ rigid filaments ($L_p/L = 1000$). The expected value of the MSD for a rigid filament of length $L$ and diameter $\sigma$ is given by $\langle \big(\mathbf R(t) - \mathbf R(0)\big)^2 \rangle = 6 D_{tr} t$, where $\mathbf R$ is the center of mass of the filament and $D_{tr}$ is the translational diffusion coefficient $D_{tr} = \frac{\ln(L/\sigma)}{3 \pi \eta L}k_B T$, where $\eta$ is the fluid viscosity. Right, simulation and theoretical comparison of the vector correlation function (VCF) averaged over $100$ rigid filaments ($L_p/L=1000$). The time axis is in simulation units $\tau$, where $\tau$ is the average time for a sphere of diameter $\sigma$ to diffuse a distance $\sigma$. The expected value of the VCF for a rigid filament of length $L$ and diameter $\sigma$ is given by $\langle \big(\mathbf u(t) - \mathbf u(0)\big)^2 \rangle = 2 \big(1 - \exp(-2 D_{r} t)\big) $, where $\mathbf u$ is the orientation of the filament and $D_{r}$ is the rotational diffusion coefficient $D_{r} = \frac{3 \ln(L/\sigma)}{\pi \eta L^3}k_B T$. The time axes are in simulation units $\tau$, where $\tau$ is the average time for a sphere of diameter $\sigma$ to diffuse its own diameter.}
  \label{fig:sim_vs_msd}
\end{figure*}

We tested the model to ensure agreement with theory for Brownian wormlike chains. Filament diffusion was validated by measuring the mean-squared displacement (MSD) and vector correlation function (VCF) for filaments in the rigid regime ($L_p \gg L$) and matching the expected values for slender, rigid filaments~\cite{doi88} (Fig.~\ref{fig:sim_vs_msd}).

\begin{figure*}[htb] \centering
  \includegraphics[width=\textwidth]{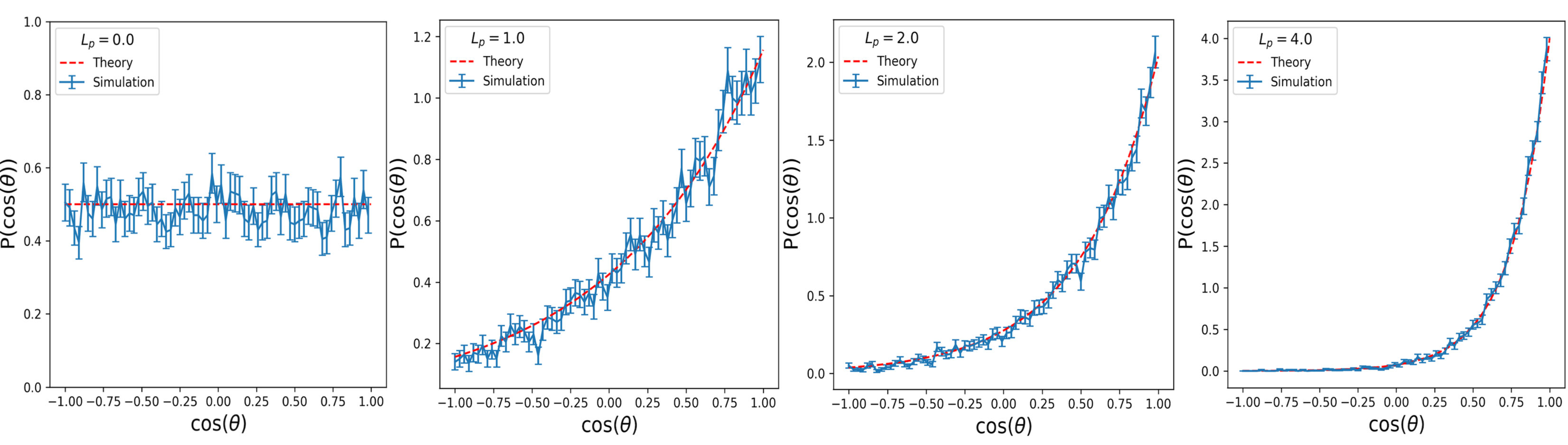}
  \caption{\small Simulation and theoretical comparison of the distributions of angles between segments for filaments with $L_p = 0$, $1$, $4$, and $8$. Simulations were carried out by averaging the results for $100$ non-interacting filaments of length $L = 20 \sigma$ and segment length $a = \sigma$ diffusing for $10^4 \tau$.  }
  \label{fig:sim_vs_boltzmann}
\end{figure*}

Filament bending was validated by ensuring that the conformations sampled by a filament at thermal equilibrium matched the expected statistical behavior. The distribution of angles between joining filament segments should be a Boltzmann distribution $P(\cos \theta) \propto e^{L_p/L \cos \theta}$. Following Montesi et al.~\cite{montesi05}, we fit a histogram of filament angles from our simulation and found good agreement with the theoretical distribution (Fig.~\ref{fig:sim_vs_boltzmann}).

\begin{figure*}[htb] \centering
  \includegraphics[width=0.7\textwidth]{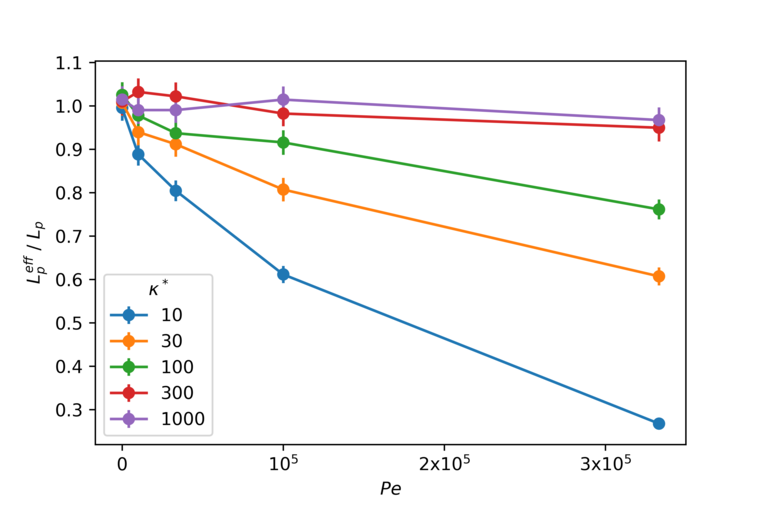}
  \caption{\small Effective persistence length derived from $\langle R^2\rangle$ for 2D wormlike chains as a function of P\'eclet number. The effective persistence lengths are plotted relative to the persistence length at zero activity. High P\'eclet numbers result in an apparent softening of the filament.}
  \label{fig:mse2e}
\end{figure*}

We validated the mean-square end-to-end distance $\langle R^2 \rangle$ of the filaments in 2D. With our choice of $\kappa = L_pk_BT$ for $d=2$, $\langle R^2 \rangle$ is given by the equation

\begin{equation}
    \langle R^2 \rangle = 4 L L_p - 8 L_p^2 (1 - e^{-\frac{L}{2 L_p} }),
\end{equation}
which differs from the usual result of a Kratky-Porod wormlike chain by the replacement $L_p\rightarrow2L_p$~\cite{kratky49}. We see an apparent softening of the filament in the presence of activity, in agreement with recent reports of the same effect~\cite{isele-holder15, singh18, gupta19, peterson20}. We plot the apparent persistence length derived from the observed $\langle R^2 \rangle$ as a function of P\'eclet number in Fig.~\ref{fig:mse2e}. The softening is appreciable at lower rigidities and vanishes for increasingly rigid filaments. 

\begin{figure*}[htb] \centering
  \includegraphics[width=0.9\textwidth]{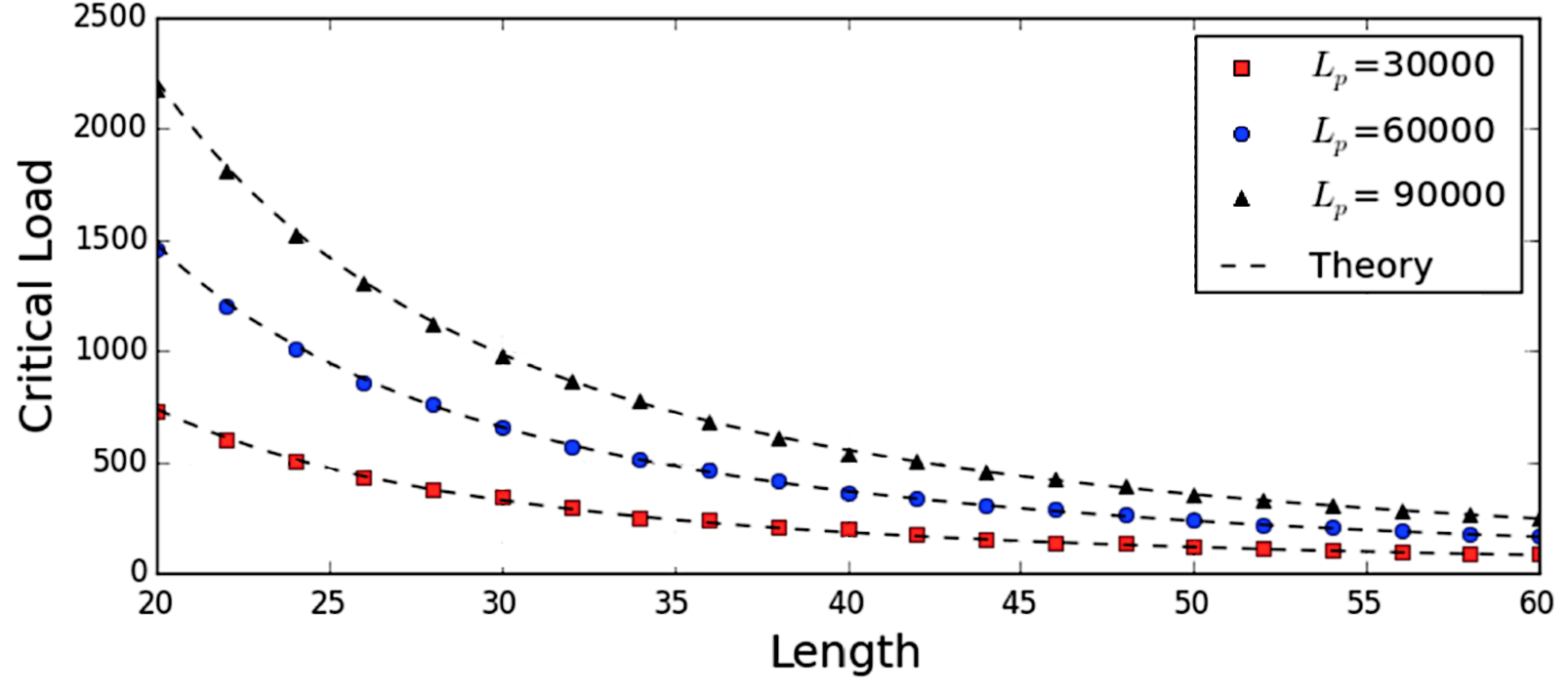}
  \caption{\small Simulation and theoretical comparison of the critical buckling load for filaments with varying length and persistence length to ensure agreement with Euler's formula for a buckling column. The critical load is given in simulation units, $k_B T / \sigma$.}
  \label{fig:sim_vs_euler}
\end{figure*}

We also validated the bending model by quantifying the filament buckling behavior for rigid filaments that were placed under a load. For a rigid filament with a persistence length $L_p$, we expect the maximum load that an unconstrained filament can withstand before buckling to be given by Euler's critical load for a column, $F_{cr} = \frac{\pi^2 L_p k_B T}{L^2}$. To measure the critical load, we linearly increased a Hookean spring force between filament ends and recorded the force at the time when the end-to-end distance of the filament sharply deviated from the filament contour length. Varying both contour length and persistence length, we compared our measured values of the critical load to theory and found good agreement with the expected values (Fig.~\ref{fig:sim_vs_euler}).

\section{Simulation parameters}
 
Key parameters of our simulation include the filament length $L$, diameter $\sigma$, persistence length $L_p$, driving force per unit length $f_{dr}$, repulsive energy $\epsilon$, simulation box diameter $L_{sys}$, and filament packing fraction $\phi$. In our simulations, all filaments have an aspect ratio $l = L/\sigma = 60$, and the system size is $l_{sys} = L_{sys}/L = 20$. In our dimensionless reduced units, $\sigma$ , $k_B T$, and $D$ are set to unity, where $D$ is the diffusion coefficient of a sphere with diameter $\sigma$. The driving force in reduced units is $f_{dr} = 3$, $10$, and $30$, so that the corresponding P\'eclet numbers are $\text{Pe} = f_{dr}L^2/k_B T \approx 10^4$, $3.3\times 10^4$, $10^5$. When the repulsive energy $\epsilon$ in the GEM-8 potential has a value $\epsilon_{dr} = 0.287 \sigma f_{dr}$, the repulsive interaction between particles induces a maximum force equal to the driving force. The repulsion parameter is then rescaled to be $\tilde{\epsilon} = \epsilon/\epsilon_{dr}$ so that $\tilde{\epsilon} = l = 60$ corresponds to fully impenetrable filaments in the absence of any additional forces for any given P\'eclet number.

The dimensionless parameters of interest when exploring the phase behavior of collective semiflexible filaments are $\tilde{\kappa} = L_p/L$, the rescaled energy $\tilde{\epsilon} = \epsilon/\epsilon_{dr}$, and packing fraction $\phi = A_\text{fil}/A_\text{sys}$, where $A_{\text{sys}}$ is the area of the simulation box and $A_{\text{fil}}=N(L\sigma + \pi\sigma^2)$ is the total area occupied by $N$ spherocylindrical filaments. The timestep used in our half step integration algorithm was $\Delta t = 10^{-4} \tau$, where $\tau$ is the average time for a sphere of diameter $\sigma$ to diffuse its own diameter. The active timescale is the time required for a filament to traverse its own length $\tau_A = l / v_{dr} = 1 / \zeta_\parallel f_{dr}$, which is $3 \tau$, $\tau$, $0.33 \tau$ for $\text{Pe} = 10^4, 3.33 \times 10^4, 10^5$ respectively.

Filaments in the simulation were initialized by randomly inserting filaments parallel to one axis of the simulation box in a nematic arrangement, and allowing the filaments to diffuse for $100 \tau$ steps before driving the filaments. Simulations terminated once they were determined to have reached a steady state, when order parameters appeared to converge to constant values. 

\section{Order parameters}
Six global order parameters quantify the system phase behavior, including polar order $P$, nematic order $Q$, average contact number $c$, average local polar order $p$, average spiral number $s$, and number fluctuations $\Delta N$. In addition, we quantified the dynamical flocking behavior of the system by characterizing the fraction of flocking filaments $N_F/N$ as well as the frequencies that filaments joined or left the flocking state, $f_\text{NF--F}$ and $f_\text{F--NF}$ respectively, which are both normalized by the number of filaments in the initial state. All order parameters are time averaged over the final 10\% of the simulation.

The polar order $P$ is the normalized magnitude of the total orientation vector of all filament segments,
\begin{equation}
  P = |\mathbf{P}| = \frac{1}{Nn} \sum_{i=1}^{Nn} \mathbf{u}_i,
\end{equation}
where $\mathbf{u}_i$ is the orientation of the $i^{\text{th}}$ filament segment for $N$ filaments each composed of $n$ segments. The polar order varies from $0$, where filaments have fully isotropic directional arrangement, to $1$, where all filaments are aligned in the same direction.

The nematic order is the maximum eigenvalue $Q$ of the 2D nematic order tensor
\begin{equation}
  \mathbf{Q}=\frac{1}{Nn}\sum_{i=1}^{Nn} (2\mathbf{u}_i\otimes\mathbf{u}_j - \mathbf{I}),
\end{equation}
where $\mathbf{I}$ is the unit tensor. The nematic order varies from $0$, with fully isotropic directional arrangement, to $1$ where all filaments are parallel or antiparallel along the same axis.

The contact number and local polar order parameters follow previous work measuring the collective behavior of active polar particles~\cite{kuan15}. The contact number is a measure of crowding in the system, and is calculated on a filament segment-wise basis,
\begin{equation}
  c_i = \sum_{\substack{j\neq i \\ \text{inter}}}^{Nn} e^{-\alpha s_{ij}^2},
\end{equation}
where $s_{ij}$ is the minimum distance between segments $i$ and $j$, and with the sum excluding all intrafilament segments. The parameter $\alpha$ determines the effective cutoff for interparticle distances, which we choose to be $1/\sigma^2$ to only consider the contributions from nearby particles to the sum. The contact number is a purely positive quantity, ranging approximately from $0$--$10$. The average contact number is the system average of the segment contact number.

The degree of local polar ordering is determined by measuring the polar order of each segment relative to their nearest neighbors, and is weighted by the segment contact number,
\begin{equation}
  \label{eqn:lpo}
  p_i = \frac{\sum_{\substack{j\neq i \\ \text{inter}}}^{Nn} \mathbf{u}_i \cdot \mathbf{u}_j e^{-\alpha s_{ij}^2}}{c_i},
\end{equation}
where the sum again excludes intrafilament segments. The local polar order ranges from $-1$, where a segment is surrounded by filaments of opposite polarity, to $1$, where a segment is surrounded by neighboring segments with the same polarity. The average local polar order is the system average of the local polar order of all filament segments.

The spiral number is a measure of filament spiraling, and is calculated by measuring the angle $\theta_i$ swept by traversing the contour of filament $i$ from tail to head originating from the center of curvature of the filament. The average spiral number is the system average
\begin{equation}
s = \frac{1}{2\pi}\sum_{i}^N \theta_i.
\end{equation}
A filament with segments that are on average oriented in a straight line will have $s\approx 0$, since the center of curvature is at a distance $\infty$ from the filament, and a filament bent into a perfect circle has $s=1$. Since filaments are not bent into perfect circles when they form a spiral, a filament spiral can be stable with a spiral number as low as $s_i\approx0.8$. Filaments may also be wound very tightly and have a spiral number $s > 1$.

\begin{figure*}[tb] \centering
  \includegraphics[width=0.66\textwidth]{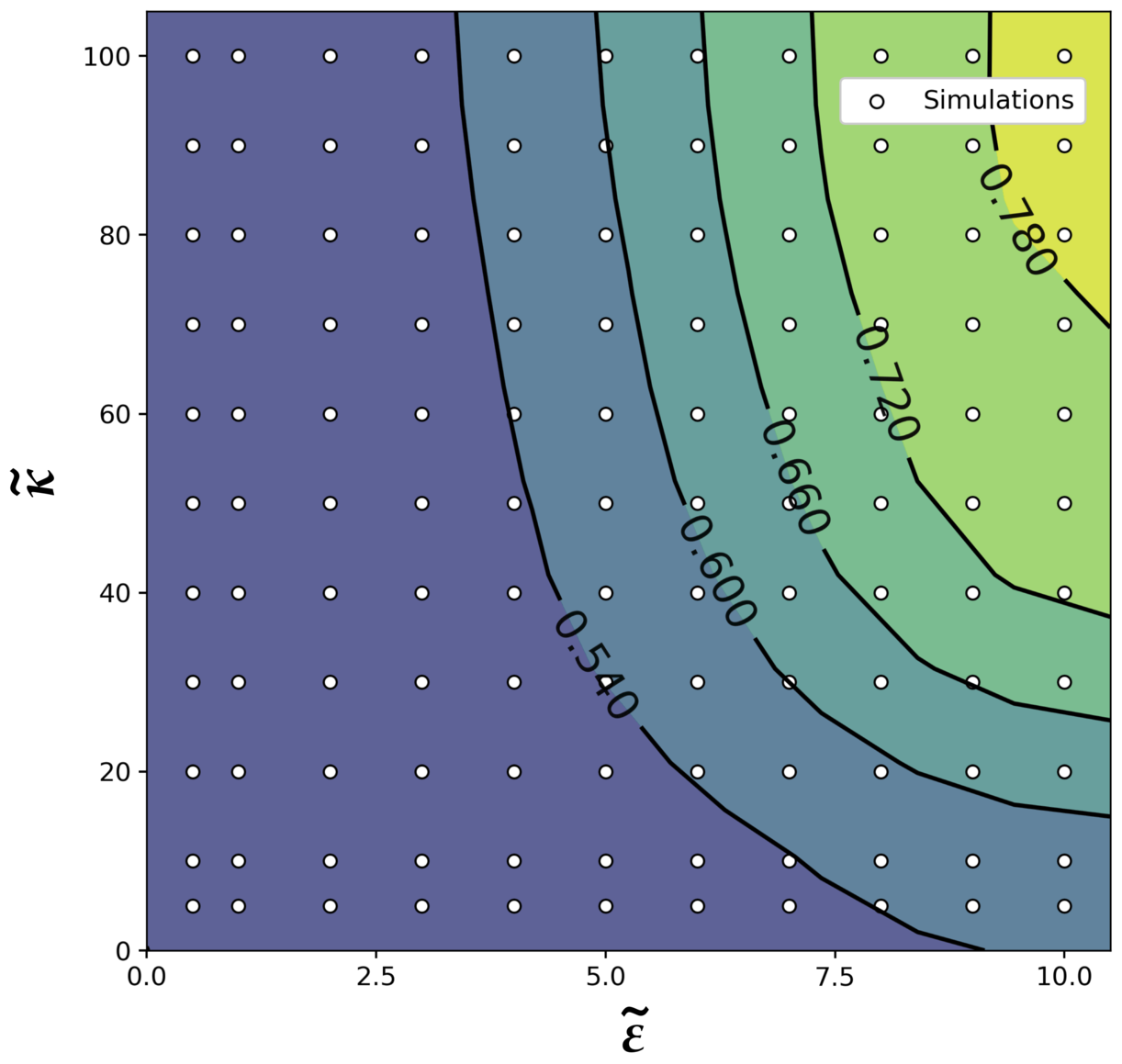}
  \caption{\small Results for the number fluctuations exponent $\alpha$, with number fluctuations $\Delta N \propto \langle N \rangle^\alpha$ for parameters $\phi=0.05$, $\text{Pe}=3.3\times 10^4$. The region of the phase diagram associated with the active isotropic phase has an exponent $\alpha\approx0.5$, whereas the regions with stable giant flocks have an exponent closer to $0.8$, the theoretical value for flocking 2D systems with long-range order.}
  \label{fig:gnf}
\end{figure*}

Number fluctuations $\Delta N$ are a measure of density fluctuations in the system. The number fluctuations are determined by drawing a box of size $L_\text{box} < L_\text{sys}$ and observing the number of filaments in the box at time $t$. By time averaging the filament number within the box, one can arrive at a mean $\langle N \rangle$ and standard deviation $\Delta N$, which is a measure of the number fluctuations. By measuring $\Delta N$ and $\langle N \rangle$ for progressively larger box sizes, one can measure the scaling of the number fluctuations relative to $\langle N \rangle$, $\Delta N \propto \langle N \rangle^\alpha$. For equilibrium systems with particles positioned at random, the central limit theorem guarantees that $\alpha=0.5$. However, systems with collective behavior have been shown to exhibit ``giant number fluctuations" (GNF), with $\alpha > 0.5$. Flocking systems that have long-range order in 2D, such as the Vicsek model at low temperature, have $\alpha = 0.8$~\cite{vicsek95, toner98, ginelli16}. We use the scaling of the number fluctuations $\alpha$ as an order parameter to determine the long-range ordering of the system, and find values of $\alpha$ that range between $0.5$--$0.8$ (Fig.~\ref{fig:gnf}).
 
We categorized the flocking behavior of filaments using the individual local order parameters $p_i$ and contact numbers $c_i$ for filaments. Following previous work~\cite{kuan15}, we determine a filament to be flocking if $p_i \geq 0.5$. We also distinguish between filaments at the flock interior and exterior by labeling flocking filaments with $c_i \geq 0.5$ as interior flocking filaments and $c_i < 0.5$ as exterior flocking filaments. We then tracked transitions between the three states, not flocking (NF), interior flocking (IF), and exterior flocking (EF) in order to determine the switching rates.

Only the total fraction of flocking filaments and the ratio of switching frequencies between the not-flocking state and flocking state were used as order parameters for clustering the simulation results. However, we found that saturation of the number of flocking filaments (giant flock phase) occurs when a large number of filaments are in the IF state, indicating that kinetic trapping of filaments plays an important role in stabilizing giant flocks, as observed in previous work~\cite{kuan15}.

\section{High-dimensional clustering of order parameters}

In order to label the simulation phases, simulation order parameters were grouped by P\'eclet number and packing fraction $\phi$ and clustered using the k-means clustering algorithm as implemented by the scikit-learn Python package~\cite{scikit-learn}. Initially, all order parameters values were standardized by $x - \mu / \sigma$, where $x$ is the order parameter value for a simulation, and $\mu$ and $\sigma$ are the mean and standard deviation of the same order parameter over all simulations with the same P\'eclet number and packing fraction. The scaled order parameters were then processed using principal component analysis, and all but the six largest components were discarded. These values were then clustered using k-means clustering for different numbers of clusters $k$, until increasing the number of clusters no longer significantly reduced the overall cost function of the algorithm. After simulations were assigned to clusters, the final simulation state behaviors were observed and labeled.

\section{Filament alignment probability}

\begin{figure*}[tb!] \centering
  \includegraphics[width=0.8\textwidth]{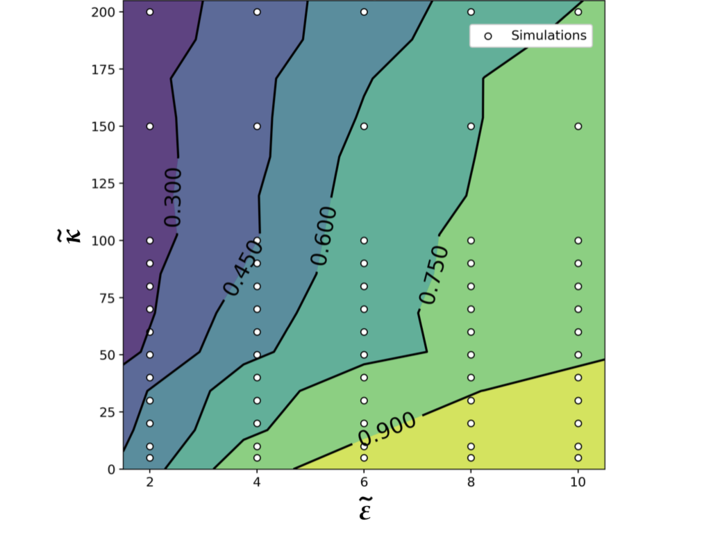}
  \caption{\small Contour plot of the probability for two randomly-oriented filaments to align upon collision ($P_\text{align}$) for different values of repulsivity and stiffness. Results were determined by averaging over 10 simulations for 100 initial collision angles $\theta_c$ for each data point. Each simulation is initialized with one filament fixed along the y-axis and the second filament oriented so that the tip is pointed at the midpoint of the first filament, and the angle between filaments is $\theta_c = \arccos(\mathbf{u}_0 \cdot \mathbf{u}_1)$ and the minimum distance between filaments is $2\sigma$. Filaments were subject to random forces, and were driven with $\text{Pe}=10$.}
  \label{fig:palign}
\end{figure*}

In order to determine the probability that two intersecting filaments align upon collision $P_{\text{align}}$, we simulated $10$ intersections of two filaments for $100$ initial collision angles evenly ranging between $0$ and $\pi$ in the presence of Brownian noise. We repeated this process while varying $\tilde{\kappa}$ and $\tilde{\epsilon}$ in order to determine the overall probability that two randomly-oriented filaments would align upon collision for a given filament stiffness and repulsivity (Fig.~\ref{fig:palign}).

\section{Filament directional persistence}

\begin{figure*}[tb] \centering
  \includegraphics[width=\textwidth]{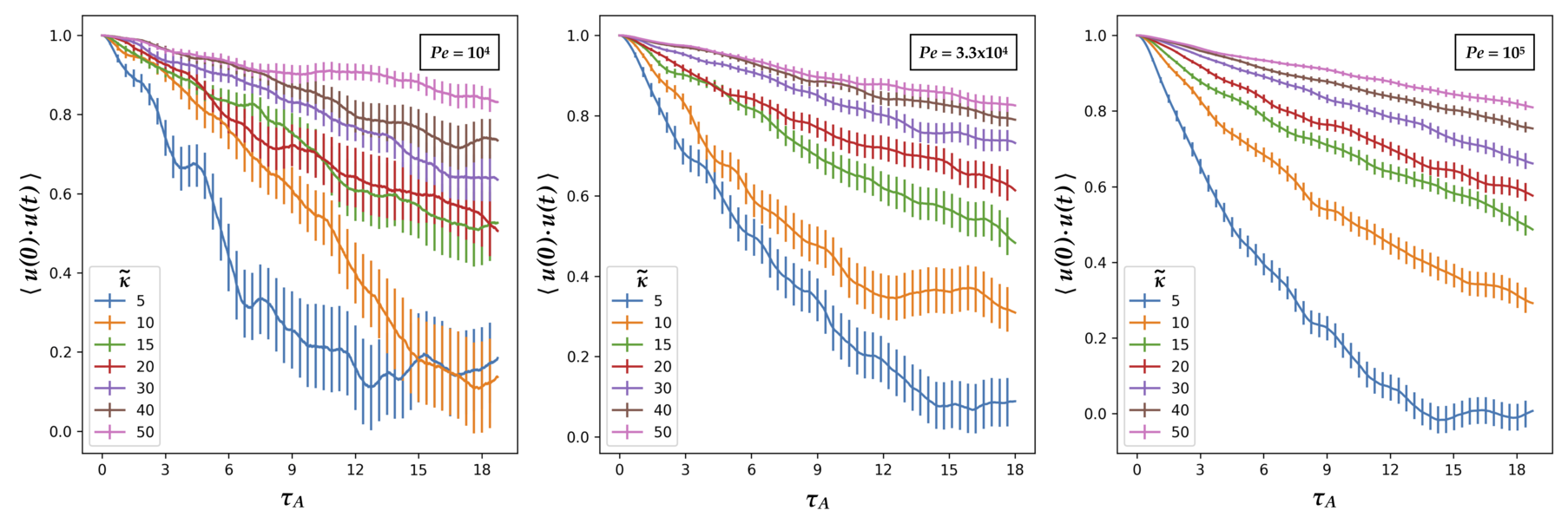}
  \caption{\small Time autocorrelation of the filament orientation $\mathbf{u}$ calculated from simulations of non-interacting filaments with polar driving force. The decorrelation time increases with increasing filament stiffness $\tilde\kappa$, indicating flexible filaments have less directional persistence than stiff filaments. Values are averaged over filament number and time, and error bars correspond to the standard error of the mean. The filaments are subject to random forces, and thus results from simulations at lower P\'eclet numbers have larger variance. Simulations were run for $10^3\tau$.}
  \label{fig:ocorr}
\end{figure*}

The filament orientation autocorrelation was calculated using
\begin{equation}
    \phi(t) = \frac{1}{T}\int_{0}^{T} \mathbf u(t') \cdot \mathbf u(t' + t) d t',
\end{equation}
where $\mathbf u(t)$ is the orientation of the filament at time $t$, and the correlation was averaged over $N$ intervals of duration $T+t$ (Fig.~\ref{fig:ocorr}). The filament orientation is determined by averaging over the orientation of filament segments.

Filament orientation autocorrelation lifetimes lengthen with increasing filament rigidity $\tilde{\kappa}$. For a fixed repulsion $\tilde{\epsilon}$, a longer orientation autocorrelation lifetime correlates with collective motion, which suggests that the angular persistence of filament trajectories contributes to the formation of persistent flocks and bands. For more ballistic trajectories, filaments that align can remain aligned for longer time, thus increasing the probability of aligning with more filaments along that trajectory, which may cause the accumulation of aligned filaments to form persistent flocks.

\section{High density simulations}

\begin{figure*}[p!] \centering
  \includegraphics[width=\textwidth]{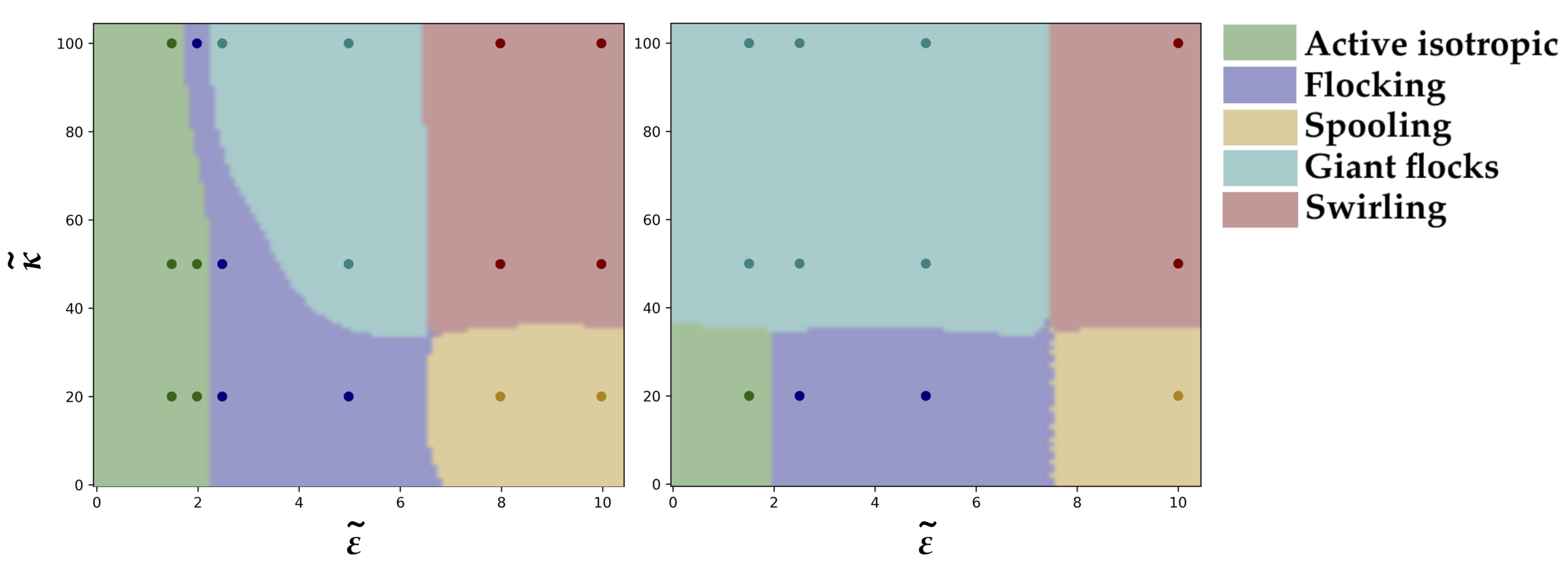}
  \caption{\small Phase diagrams for simulations at higher filament densities with P\'eclet number $\text{Pe}=10^5$. Filament packing fractions are $\phi=0.2$ (left), and $\phi=0.4$ (right).}
  \label{fig:highphi}
\end{figure*}

\begin{figure*}[htb] \centering
  \includegraphics[width=0.6\textwidth]{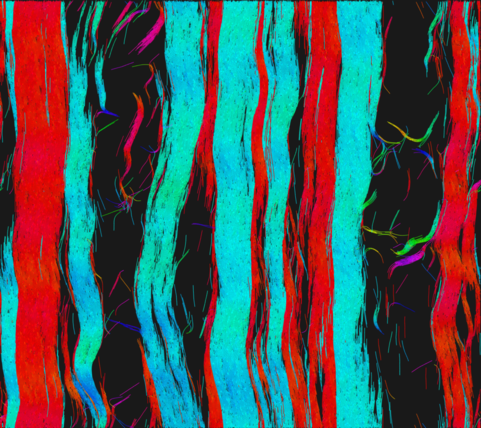}
  \caption{\small Simulation with nematic laning bands for parameters $\phi=0.8$, $\tilde\kappa=100$, $\tilde\epsilon=10$, $\text{Pe}=10^5$.}
  \label{fig:pf0.8}
\end{figure*}

Simulations were run at $\phi=0.2$ and $0.4$, for $\tilde\kappa=20$, $50$, and $100$, and for values of $\tilde\epsilon$ ranging from $1.5$--$10$ in order to explore the parameter space at higher filament densities and search for density-dependent phase behavior. We find that the giant flock phase becomes more stable at a larger range of values of $\tilde\epsilon$ with increasing density (Fig.~\ref{fig:highphi}). At $\phi=0.4$, we found that multiple polar bands can coexist in the giant flocking phase, resulting in nematic laning, and is stable over a wide range of repulsivity. We also ran one simulation at $\phi=0.8$ at $\tilde\epsilon=10$ and $\tilde\kappa=100$, and find the giant flock/nematic laning phase to be stable at higher values of $\tilde\epsilon$ (Fig.~\ref{fig:pf0.8}). 

We labeled the results by comparing the high density order parameter values to the results from simulations at lower filament densities. The only major inconsistency in the order parameter values between the labeled phases for low and high densities is the global polar order parameter $P$, which is high in the giant flock phase at densities $\phi \leq 0.2$, and is closer to zero at higher densities due to the presence of nematic laning bands. However, we classify nematic laning as a type of giant flocking phase, since the underlying dynamical behavior is similar.

\section{Intrinsic curvature}

An intrinsic curvature was added to the filament model by modifying the bending potential in Eqn.~\ref{ubend} to have an offset angle $\phi_0$,

\begin{equation}\label{eqn:ubendic}
  U_{\text{bend}} = - \frac{\kappa}{a}\sum_{k=2}^{N-1} \cos{(\theta_{k,k-1} - \phi_0)},
\end{equation}

where $\theta_{k, k-1} = \arccos{(\mathbf{u}_k\cdot\mathbf{u}_{k-1})}$ is the angle between site orientations $k$ and $k-1$, and $\phi_0=ad\phi/ds$ corresponds to the expected angle between two segments of length $a$ with a curvature per unit length $d\phi/ds$.

It can be shown that the term in the sum of Eqn.~\ref{eqn:ubendic} can be rewritten as
\begin{equation}
    \cos{(\theta_{k,k-1} - \phi_0)} = \mathbf{R} \mathbf u_{k} \cdot \mathbf{R}^{-1} \mathbf u_{k-1},
\end{equation}
where $\mathbf R$ is a rotation matrix that rotates the orientation vector $\mathbf u_k$ by an angle $\phi_0/2$,
\begin{equation}
    \mathbf R = 
        \begin{pmatrix}
            \cos(\phi_0/2) & -\sin(\phi_0/2) \\
            \sin(\phi_0/2) & \cos(\phi_0/2)
        \end{pmatrix},
\end{equation}
and its inverse $\mathbf R^{-1}$ rotates the orientation vector $\mathbf u_{k-1}$ by an angle $-\phi_0/2$. The combined bending and metric forces from Eqn.~\ref{eqn:fbendfmetric} with intrinsic curvature are therefore
\begin{equation}
  \mathbf{F}_i^{\text{bend}}+\mathbf{F}_i^{\text{metric}} = \frac{1}{a}\sum_{k=2}^{N-1}\kappa_k^{\text{eff}}\frac{\partial (\mathbf R \mathbf{u}_k\cdot\mathbf R^{-1} \mathbf{u}_{k-1})}{\partial \mathbf{r}_i},
\end{equation}
which can be expanded in the same way as Eqn.~\ref{eqn:fbendfmetricx}.

\begin{figure*}[p!] \centering
  \includegraphics[width=0.6\textwidth]{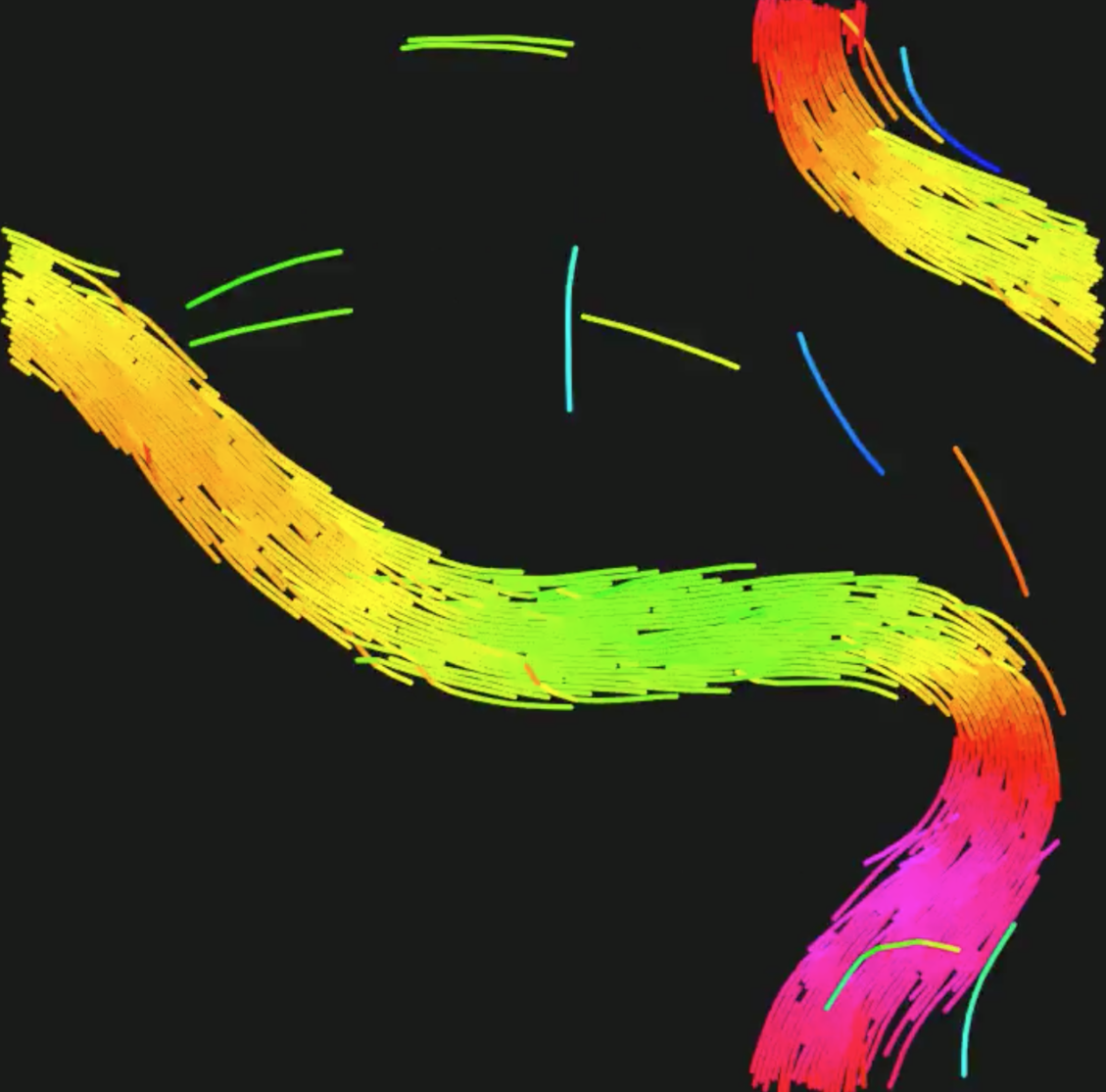}
  \caption{\small Movie of simulation with intrinsic curvature $0.02$ rad/$\sigma$ and parameters $\phi=0.3$, $\tilde{\kappa}=100$, $\tilde{\epsilon}=5$, $\text{Pe}=10^5$.}
  \label{mov:intrinsic_curvature}
\end{figure*}

We ran a simulation with intrinsic curvature $d\phi/ds = 0.02$ radians/$\sigma$, with packing fraction $\phi = 0.3$, filament length $l = 37$, system box length $l_{sys} = 6.67 l$, rigidity $\tilde{\kappa} = 100$, and repulsion $\tilde{\epsilon} = 5$ (Fig.~\ref{mov:intrinsic_curvature}). The filaments coalesce into a single polar band, which then buckles and reassembles at an angle rotated in the direction of filament curvature. This buckling and rotation behavior continues for the length of the simulation. The rotation of the polar order vector has been observed in filament gliding assay experiments and simulations of bead-spring filament models with intrinsic curvature~\cite{kim18}.

\bibliographystyle{unsrt} \bibliography{cmdsf_supplement} 

\clearpage

\clearpage

\begin{figure*}[p!] \centering
  \includegraphics[width=0.6\textwidth]{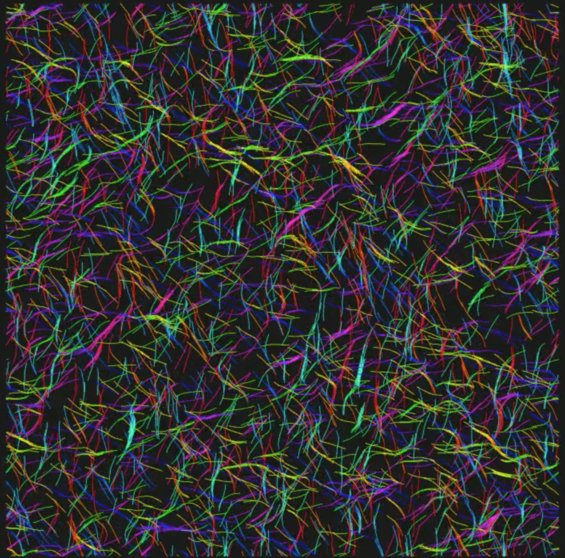}
  \caption{\small Movie of simulation with filaments in the active isotropic phase with parameters $\phi=0.2$, $\tilde{\kappa}=20$, $\tilde{\epsilon}=1.5$, $\text{Pe}=10^5$.}
  \label{mov:active_iso}
\end{figure*} \clearpage

\begin{figure*}[htb] \centering
  \includegraphics[width=0.6\textwidth]{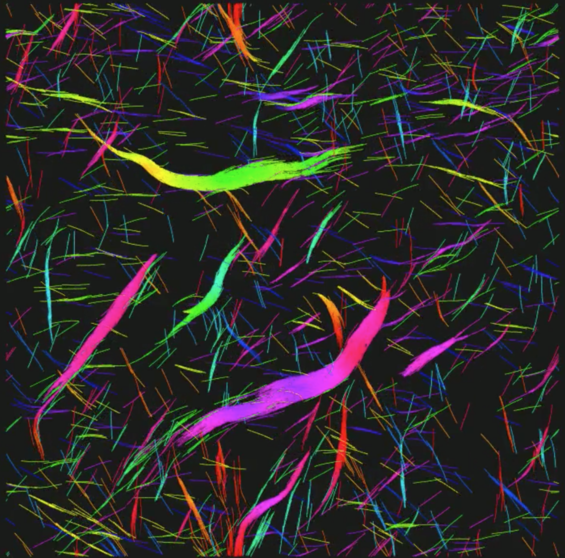}
  \caption{\small Movie of simulation with filaments in the flocking phase with parameters $\phi=0.2$, $\tilde{\kappa}=100$, $\tilde{\epsilon}=2$, $\text{Pe}=10^5$.}
  \label{mov:flocking}
\end{figure*}\clearpage

\begin{figure*}[p!] \centering
  \includegraphics[width=0.6\textwidth]{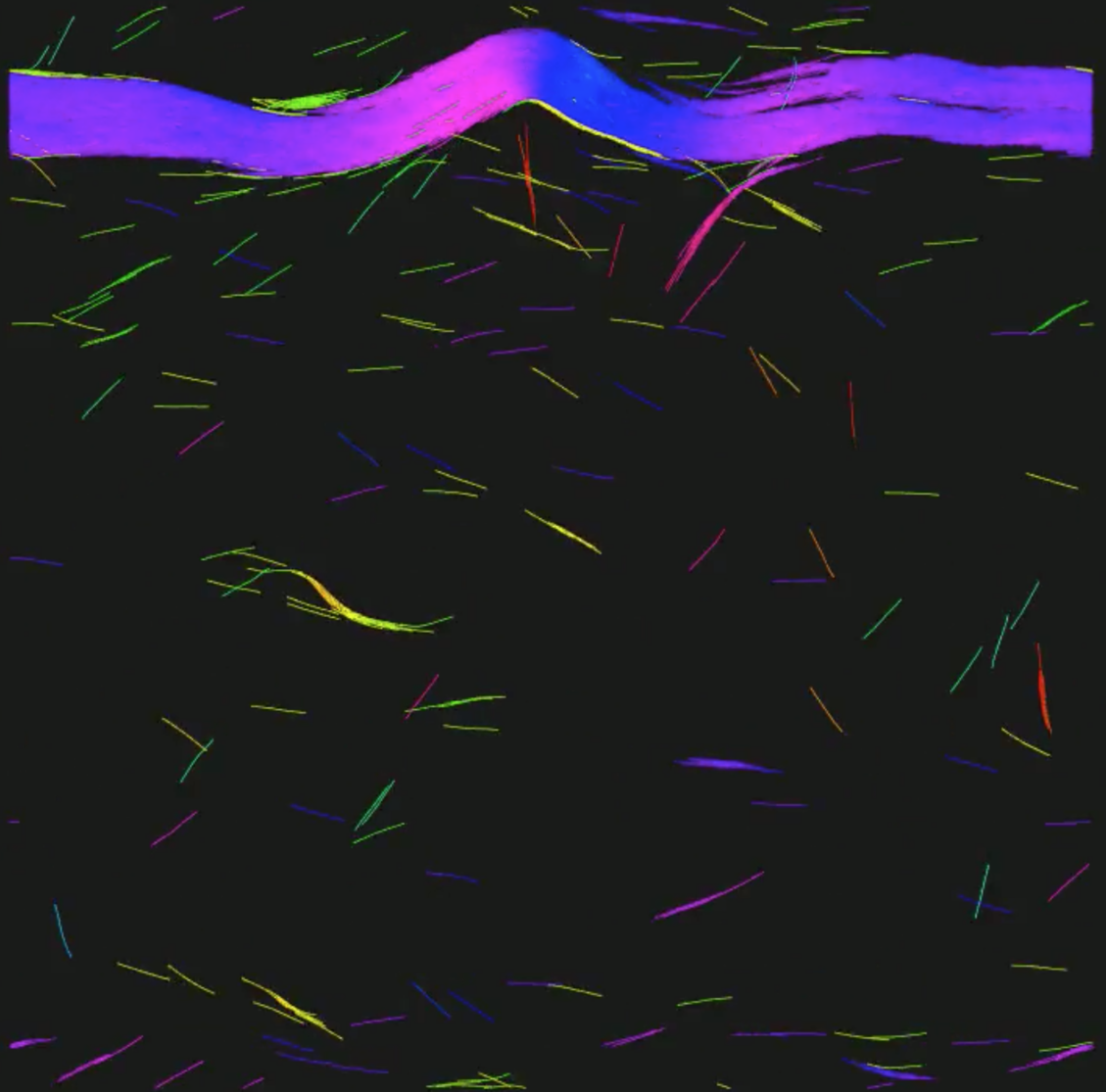}
  \caption{\small Movie of simulation with filaments in the polar band phase with parameters $\phi = 0.2$, $\tilde{\kappa}=100$, $\tilde{\epsilon}=3$, $\text{Pe}=10^5$.}
  \label{mov:polar_band}
\end{figure*}\clearpage

\begin{figure*}[htb] \centering
  \includegraphics[width=0.6\textwidth]{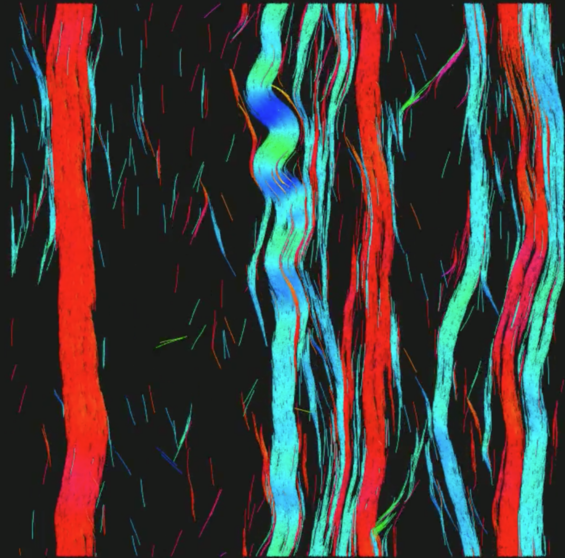}
  \caption{\small Movie of simulation with filaments in the nematic band phase with parameters $\phi=0.4$, $\tilde{\kappa}=100$, $\tilde{\epsilon}=5$, $\text{Pe}=10^5$.}
  \label{mov:nematic_band}
\end{figure*}\clearpage

\begin{figure*}[p!] \centering
  \includegraphics[width=0.6\textwidth]{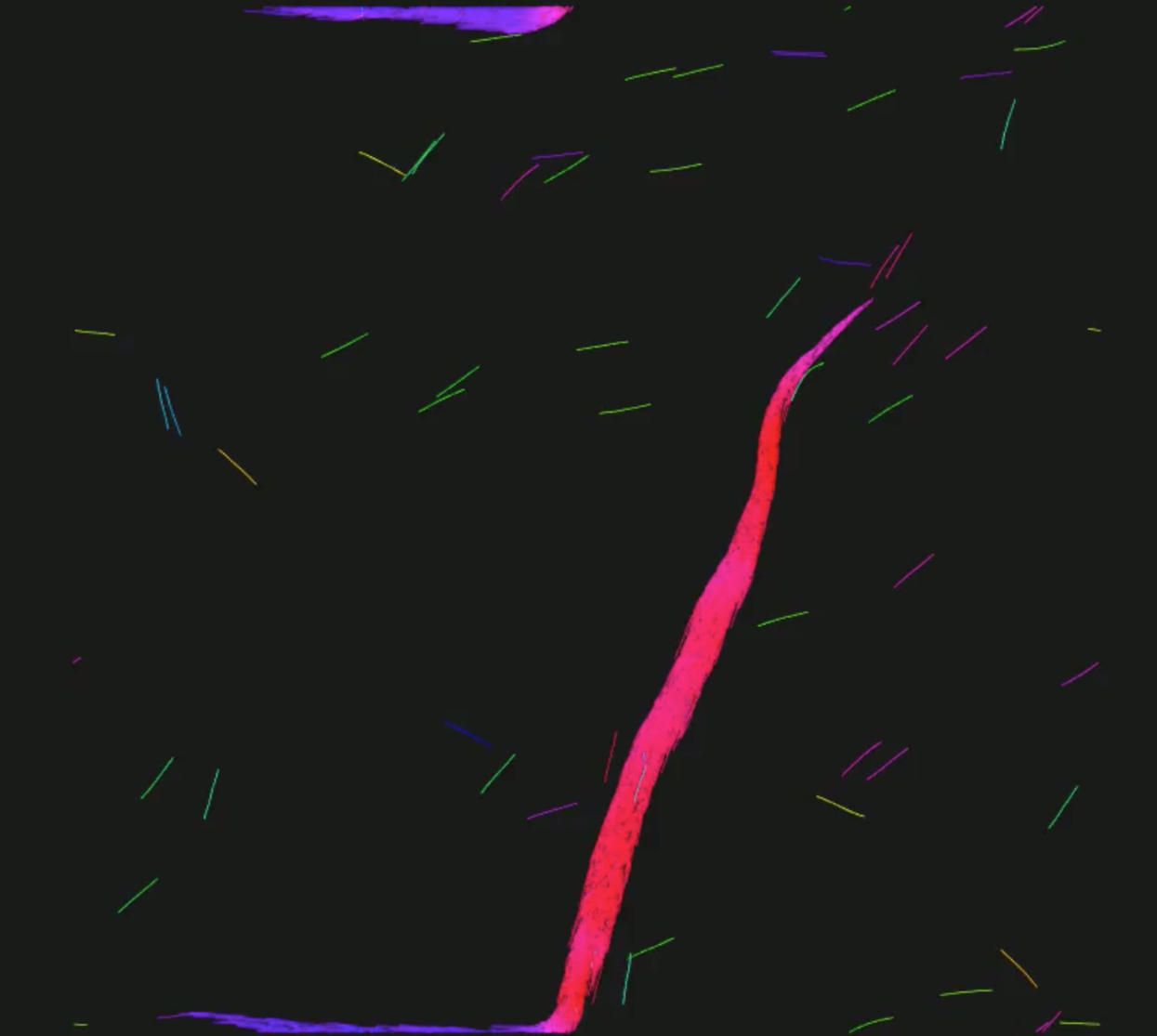}
  \caption{\small Movie of simulation with filaments in the giant flock phase with parameters $\phi=0.04$, $\tilde{\kappa} = 100$, $\tilde{\epsilon} = 5$, $\text{Pe}=10^5$.}
  \label{mov:giant_flock}
\end{figure*}\clearpage

\begin{figure*}[htb] \centering
  \includegraphics[width=0.6\textwidth]{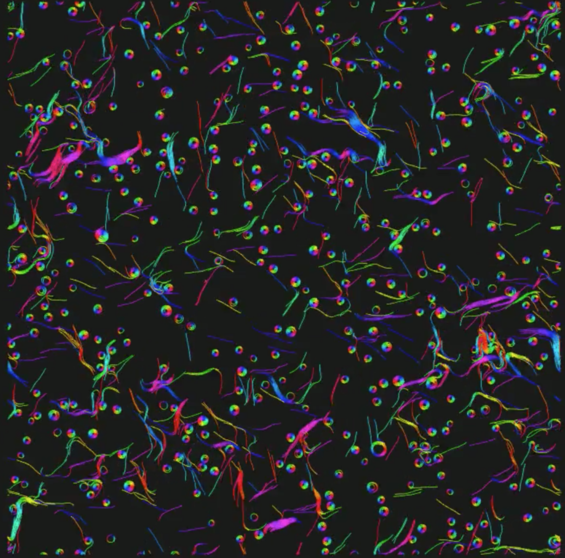}
  \caption{\small Movie of simulation with filaments in the spooling phase with parameters $\phi=0.1$, $\tilde{\kappa}=20$, $\tilde{\epsilon}=10$, $\text{Pe}=10^5$.}
  \label{mov:spooling}
\end{figure*}\clearpage

\begin{figure*}[p!] \centering
  \includegraphics[width=0.6\textwidth]{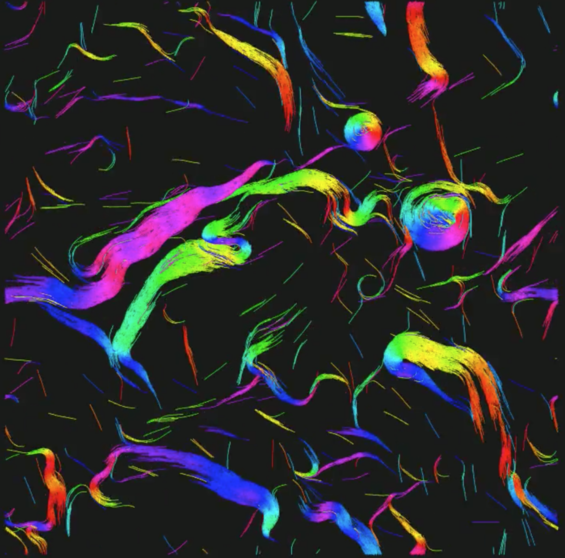}
  \caption{\small Movie of simulation with filaments in the swirling phase with parameters $\phi=0.2$, $\tilde{\kappa}=100$, $\tilde{\epsilon}=10$, $\text{Pe}=10^5$.}
  \label{mov:swirling}
\end{figure*}\clearpage

